\documentclass{aa}
\usepackage{graphicx}
\usepackage{txfonts}
\graphicspath{ {figures/} }

\begin{document} 

\title{The impact of third dredge-up on the mass loss of Mira variables\thanks{The reduced spectra are only available in electronic form at the CDS via anonymous ftp to cdsarc.u-strasbg.fr (130.79.128.5) or via http://cdsweb.u-strasbg.fr/cgi-bin/qcat?J/A+A/.}}

\author{
S.~Uttenthaler\inst{\ref{inst_iap}}\fnmsep\thanks{Corresponding author: S. Uttenthaler,\\ {\tt stefan.uttenthaler@gmail.com}} 
\and
S.~Shetye\inst{\ref{KUL}}
\and
A.~Nanni\inst{\ref{AN1},\ref{AN2}}
\and
B.~Aringer\inst{\ref{UV}}
\and
K.~Eriksson\inst{\ref{UU}}
\and
I.~McDonald\inst{\ref{iain1},\ref{iain2}}
\and 
D.~Gobrecht\inst{\ref{KUL}}
\and
S.~H\"ofner\inst{\ref{UU}}
\and
U.~Wolter\inst{\ref{inst_hamburg}}
\and
S.~Cristallo\inst{\ref{AN2},\ref{sergio2}}
\and
K.~Bernhard\inst{\ref{inst_aavso}}
}


\institute{
Institute of Applied Physics, TU Wien, Wiedner Hauptstra\ss e 8-10, 1040 Vienna, Austria\label{inst_iap}
\and
Instituut voor Sterrenkunde, KU Leuven, Celestijnenlaan 200D, bus 2401, 3001, Leuven, Belgium\label{KUL}
\and
National Centre for Nuclear Research, ul. Pasteura 7, 02-093 Warsaw, Poland\label{AN1}
\and
INAF - Osservatorio astronomico d’Abruzzo, Via Mentore Maggini, snc, 64100 Teramo, Italy\label{AN2}
\and
Department of Astrophysics, University of Vienna, T\"urkenschanzstra\ss e 17, 1180 Vienna, Austria\label{UV}
\and
Theoretical Astrophysics, Department of Physics and Astronomy, Uppsala University, Box 516, 75120, Uppsala, Sweden\label{UU}
\and
Jodrell Bank Centre for Astrophysics, Department of Physics and Astronomy, School of Natural Sciences, University of Manchester, M13 9PL Manchester, UK\label{iain1}
\and
Department of Physical Sciences, The Open University, Walton Hall, MK7 6AA Milton Keynes, UK\label{iain2}
\and
Hamburger Sternwarte, Universit\"at Hamburg, Gojenbergsweg 112, 21029 Hamburg, Germany\label{inst_hamburg}
\and
INFN, Sezione di Perugia, Via A. Pascoli, snc, 06123 Perugia, Italy\label{sergio2}
\and
American Association of Variable Star Observers (AAVSO), Cambridge, MA, USA\label{inst_aavso}
}

\date{Received July 29, 2024; accepted September 05, 2024}


\abstract
{The details of the mass-loss process in the late stages of low- and intermediate-mass stellar evolution are not well understood, in particular its dependence on stellar parameters. Mira variables are highly suitable targets for studying this mass-loss process.}
{Here, we follow up on our earlier finding that a near-to-mid-infrared (NIR-MIR) colour versus\ pulsation period diagram shows two sequences of Miras that can be distinguished by the third dredge-up (3DUP) indicator technetium in those stars. While IR colours are good indicators of the dust mass-loss rate (MLR) from Miras, no corresponding sequences have been found using the gas MLR. However, investigations of the gas MLR have been hampered by data limitations. We aim to alleviate these limitations with new observational data.}
{We present new optical spectra of a well-selected sample of Miras. We searched these spectra for absorption lines of Tc and other 3DUP indicators, and combine our findings with gas MLRs and expansion velocities from the literature. Furthermore, we extend the analysis of the MIR emission to WISE data and compare the broadband spectral energy distributions (SEDs) of Miras with and without Tc.}
{We find no systematic difference in gas MLRs between Miras with and without Tc. However, the gas envelopes of Tc-poor Miras appear to have a higher terminal expansion velocity than those of Miras with Tc. Furthermore, our analysis of the IR photometry strongly corroborates the earlier finding that Tc-poor Miras have a higher MIR emission than Tc-rich ones, by as much as a factor of two. We model the IR colours with DARWIN and stationary wind models and conclude that Miras with and without Tc have different dust content or dust properties.}
{We discuss several hypotheses and interpretations of the observations and conclude that the reduction of free oxygen by 3DUP of carbon and iron-depleted dust grains in Tc-rich stars are the most convincing explanations for our observations.}

\maketitle

\section{Introduction}\label{sec:intro}

Mira variables are red giants in the asymptotic giant branch (AGB) phase of the evolution of low- to intermediate-mass ($0.8\lesssim M/M_{\sun} \lesssim 8$) stars. Three important properties characterise AGB stars: (i) large-amplitude, long-period pulsation of their atmospheres; (ii) significant mass loss through stellar winds; and, at least in the later stages of the AGB, (iii) enrichment of newly synthesised elements in their atmospheres through a deep-mixing process called the third dredge-up (3DUP).


The pulsation of the outer atmosphere leads to striking variability in the apparent brightness of the stars over timescales of $\sim100$ to over 1000\,d, with amplitudes of $\Delta V > 2\fm5$ in the visual passband \citep{2017ARep...61...80S}. This makes Miras readily detectable among other red giant stars in large-scale surveys 
\citep[e.g.][]{2016ApJS..227....6V,2021A&A...648A..44M}.

It is believed that the pulsation directly impacts the mass loss (ML) from Mira stars. Shock waves accompanying large-amplitude pulsations are responsible for the substantially higher gas density in the cooler outer layers of the stellar atmosphere, leading to efficient growth of dust grains in the supersaturated gas. The dust grains scatter and absorb light from the stellar photosphere, thus gaining momentum that pushes them away from the star. Momentum is transferred to gas atoms and molecules via collisions, thus initiating a stellar wind that eventually leaves the star. This wind mechanism has been called the pulsation-enhanced dust-driven outflow scenario \citep{2008A&A...491L...1H}. However, the relative importance of pulsation and dust for launching the wind may vary with stellar mass and evolutionary state \citep{2018MNRAS.481.4984M}. Generally, mass-loss rates (MLRs) of Miras range from $\sim10^{-8}$ to $\sim10^{-4} M_{\sun}{\rm yr}^{-1}$, and the wind creates a circumstellar envelope (CSE) around the star that is rich in molecules and dust. Eventually, ML from the star will erode the hydrogen envelope so much that it will terminate the evolution on the AGB; see \citet{2018A&ARv..26....1H} for a review of the ML process of AGB stars.

Third dredge-up events, on the other hand, occur quasi-periodically (depending on mass, every $10^4-10^5$ years) during the later stages of the AGB after vigorous ignitions of the He-burning shell called thermal pulses (TPs) or He-shell flashes. After a TP, the convective envelope of the star may reach deep layers, where nuclear burning has taken place, mixing up the burning products to the stellar surface \citep[e.g.][]{2006NuPhA.777..311S}. In the context of this paper, two isotopes are of particular importance: $^{12}$C and $^{99}$Tc.

Carbon is a direct product of He burning. Repeated dredge-up of $^{12}$C will turn the initially oxygen-rich (spectral type M, ${\rm C}/{\rm O}\approx0.5$ by number) star into a carbon-rich (spectral type C, ${\rm C}/{\rm O}>1$) star, with spectral type S being an intermediate transition type ($0.5\lesssim{\rm C}/{\rm O}<1$). This increase in carbon abundance will eventually fundamentally change the chemistry of the star's atmosphere and its spectral appearance \citep{2010JApA...31..177L}. 

Technetium (Tc), on the other hand, is an element with only radioactively unstable isotopes. Its longest-lived isotope produced by the slow neutron capture ($s$-)process in AGB stars, $^{99}$Tc, has a half-life of $\sim2\times10^5$\,yr, which is short compared to the $\sim10^6$\,yr a star spends on the upper AGB \citep[e.g.][]{2020MNRAS.498.3283P}. $^{99}$Tc thus serves as an indicator of recent or ongoing 3DUP in an AGB star well before an enhancement of $^{12}$C abundance is detectable \citep{1952ApJ...116...21M,1987AJ.....94..981L}, and can be used to discriminate between intrinsic (i.e.\ through internal nucleosynthesis and mixing) and extrinsic (i.e.\ through external mass transfer from a former AGB binary companion) $s$-process enrichment \citep{1988ApJ...333..219S,1993A&A...271..463J}. Based on the presence of atmospheric Tc absorption lines in the blue optical range ($\sim4200-4300$\,\AA), one can easily distinguish between pre-3DUP (Tc-poor) and post-3DUP (Tc-rich) AGB stars. 
Indeed, Mira variables with and without Tc have been found from observations \citep{1987AJ.....94..981L}. These are also the two main groups of stars analysed and compared in the present paper with respect to the effect of 3DUP on the ML process.

The change in atmospheric chemistry due to 3DUP may also change essential details of the ML process. It is well established that, in C-rich stars, momentum transfer happens via the absorption of stellar photons by grains of amorphous carbon and SiC \citep[e.g.][]{2000A&A...361..641W,2003A&A...399..589H}. However, for O-rich stars, the nature of the wind-driving grains is not firmly established \citep{2006A&A...460L...9W}. In this case, it is thought that the momentum transfer is provided by the scattering of photons off the relatively large silicate grains (Mg$_2$SiO$_4$) that can form around smaller alumina (Al$_2$O$_3$) seed particles \citep{2008A&A...491L...1H,2016A&A...594A.108H,2016A&A...585A...6G}, or by a combination of absorption and scattering on iron-rich grains \citep{2002A&A...384..585K,2010ApJ...717L..92M,nanni13}. Though observational evidence suggests that the overall ML properties of C-rich and O-rich stars at solar metallicity are similar, some differences may be found in the details \citep[e.g.][]{2015A&A...581A..60D}.

Furthermore, a binary companion to the AGB star may also impact the ML properties. For example, hydrodynamical models of binary mass loss by \citet{2021A&A...653A..25M} suggest that a close stellar-mass companion may significantly increase the terminal velocity of the outflow from a mass-losing AGB star. Recent observational evidence suggests that the companions significantly shape the winds from evolved stars \citep{2020Sci...369.1497D}. It is therefore appropriate to take special care when analysing observed MLRs, and when comparing the overall ML properties of O-rich to C-rich stars, and the ML properties of (putative) single stars to those of known binaries.

\citet{2013A&A...556A..38U} and \citet[][hereafter paper~II]{2019A&A...622A.120U} reported that, at a given pulsation period, Miras with Tc have lower near-infrared (NIR) to mid-infrared (MIR) colours than Miras without Tc. We refer to Figs.~2 and 3 of paper~II to clearly demonstrate this dichotomy. NIR to MIR colours such as $K-W4$, where $K$ is the mean magnitude in the $K$-band and $W4$ is the magnitude in the 22\,$\mu{\rm m}$ band of the Wide-field Infrared Survey Explorer (WISE) space observatory, are an indicator of the dust MLR of AGB stars ($W4$ was denoted $[22]$ in paper~II). A larger $K-W4$ colour indicates a larger dust MLR. A lower dust MLR from Tc-rich, post-3DUP Miras is unexpected because these objects are thought to be more evolved, and MLR is expected to increase during the evolution on the AGB. Theoretical models do not predict a MLR decrease except for short phases during a TP \citep[e.g.][]{1993ApJ...413..641V}. A possible decrease in the dust MLR as a direct result of 3DUP was thus discussed in paper~II as one of several hypotheses to explain the observational evidence.

However, most of the mass is lost as gas, not dust. Because the gas-to-dust mass ratio is usually $>100$ \citep[e.g.][]{2008A&A...487..645R,2017MNRAS.465..403G}, the total MLR may be set equal to the gas MLR, $\dot{M}_{\rm gas}$, which is usually derived from millimeter-wave CO emission-line observations. Therefore, paper~II investigated whether or not such a dichotomy between Tc-poor and Tc-rich Miras is also present in the $\dot{M}_{\rm gas}$ versus\ $P$ diagram, but no such dichotomy was not detected. However, there were significant limitations to the available $\dot{M}_{\rm gas}$ data that prevented firm conclusions from being drawn on the possible impact of 3DUP on the gas MLR of Miras. Only 37 Miras were identified for which information on both $\dot{M}_{\rm gas}$ and the presence of Tc was available. Even fewer have pulsation periods below 350\,d, where the dichotomy in $K-W4$ is most evident. Finally, significant uncertainties could be attached to the $\dot{M}_{\rm gas}$ determination from CO lines \citep{2008A&A...487..645R}.

The aim of the present paper therefore is to improve on the available data and revisit the question of whether or not 3DUP could impact (reduce) the MLR of Mira stars. This aim can be achieved by increasing the number of Miras for which information on  both $\dot{M}_{\rm gas}$ and the presence of Tc is available. There are ongoing large observing programmes to increase the number of AGB stars, including Miras, with well-measured $\dot{M}_{\rm gas}$; see for example 
\citet{2020A&A...640A.133R,2022A&A...660A..94G,2022MNRAS.512.1091S}. For the present paper, we opted to improve the database by observing several Miras with high-resolution optical spectrographs for which a measurement of $\dot{M}_{\rm gas}$ from CO lines is already available in the literature. The Tc content of the targets, and hence their 3DUP activity, were subsequently derived from the new spectra. Here, we report the results of these observations, reanalyse the ML properties of Miras with and without Tc, and provide further evidence that the amount and/or properties of the dust around Miras depend on their 3DUP activity.


\section{Target selection and observations}\label{sec:observations}

The observations used in this paper were carried out with the Heidelberg Extended Range Optical Spectrograph (HEROS) mounted on the robotic 1.2\,m TIGRE telescope located near Guanajuato in central Mexico \citep{2014AN....335..787S}, and the High-Efficiency and high-Resolution Mercator Echelle Spectrograph (HERMES) mounted on the 1.2\,m Mercator telescope on the island of La Palma, Spain \citep{2011A&A...526A..69R}. The HEROS spectrograph provides a spectral resolving power of $R=20\,000$ from 3800 to 8800\,\AA\ with a small gap between the two arms from 5700 to 5830\,\AA. HERMES, on the other hand, provides a spectral resolving power of $R=85\,000$ in the high-resolution fibre, with spectral coverage from 3770 to 9000\,\AA\ in a single exposure. All spectra have been reduced using the respective standard reduction pipeline. The spectra are only flat-fielded and not flux-calibrated, which will be sufficiently accurate for the relatively short spectral pieces investigated here.

For the observations with TIGRE/HEROS, targets were selected from Table~A.2 of paper~II among those stars that had no prior Tc classification in the literature. As a second criterion, we selected Miras with pulsation periods in the range where the dichotomy in $K-W4$ between Tc-poor and Tc-rich Miras exists, 
that is, between 250 and 400 days of period. Due to the faintness of the targets during minimum light, we aimed to observe them close to maximum visual light. AAVSO\footnote{\url{https://www.aavso.org/LCGv2/}} light curves were used to plan optimal observing dates for each target. Some targets were observed over multiple nights. In particular, the spectra of U~Ari from 2019 had a signal-to-noise ratio (S/N) 
that was too low to be useful; thus, the observations were repeated.

Also, we aimed to have equal numbers of Tc-poor and Tc-rich Miras in the sample. Therefore, we inspected their location in the $K-W4$ vs.\ $P$ diagram with respect to the best-separating line defined in the lower panel of Fig.~3 of paper~II. The accuracy (Fraction Correctness) of this separating line is $\sim87$\%. Thus, a few incorrect predictions can be expected. Indeed, four of the targets predicted in this way to be Tc-rich (T~Ari, W~Cnc, X~Oph, R~Peg) turned out to be Tc-poor. None of the stars predicted to be Tc-poor were found to be Tc-rich.

The data from Mercator/HERMES were not observed specifically for this study, but we searched the archive for objects from Table~A.2 without previous Tc classification in the literature. We found spectra of five sample stars. These have pulsation periods over 400\,d, but they are nevertheless interesting enough to be included here. The log of all TIGRE/HEROS and Mercator/HERMES observations is presented in Table~\ref{log-obs}. The table lists the variable name, used telescope/instrument combination, observation date, exposure time, and the S/N. We estimate the S/N of the spectra on a 25\,\AA\ wide interval approximately centred on the 4262\,\AA\ line, using the noise estimates determined by the optimum extraction procedure of the TIGRE and HERMES pipelines. The S/N for our other Tc lines does not differ significantly from these values. If several exposures were taken in one night, we estimate the S/N of the combined spectrum.

\begin{table}
\caption{Observation log.}\label{log-obs}
\centering
\begin{tabular}{lccrr}
\hline\hline
Name   & Inst. & Obs.\ date    & Exp. time & S/N \\
       &       & yyyy/mm/dd    & s         &     \\
\hline
T~Ari  & T & 2019/08/13 & 1800             &  26 \\
U~Ari  & T & 2019/02/21 & $3\times400$     &  10 \\
       & T & 2019/02/27 & 1200             &  10 \\
       & T & 2020/01/07 & 1500             &  37 \\
       & T & 2020/01/13 & 1500             &  34 \\
R~Aur  & M & 2013/03/28 & 100              & 100 \\
U~Aur  & M & 2012/09/24 & 400              &  18 \\
R~Cas  & M & 2011/10/22 & 100              & 100 \\
WY~Cas & M & 2011/09/27 & 350              &  30 \\
S~CMi  & T & 2019/02/21 & $3\times400$     &  21 \\
W~Cnc  & T & 2019/10/06 & 1800             &  22 \\
U~Her  & M & 2016/04/08 & 960              & 300 \\
X~Hya  & T & 2019/02/21 & $3\times600$     &  11 \\
R~Leo  & T & 2019/02/03 & 600              &  62 \\
       & T & 2019/02/18 & 600              &  55 \\
R~LMi  & T & 2019/05/11 & 540              &  25 \\
       & T & 2019/05/17 & 540              &  23 \\
       & T & 2019/05/22 & 540              &  32 \\
R~Lyn  & T & 2019/09/15 & 1800             &  15 \\
X~Oph  & T & 2019/03/04 & 900              &  25 \\
R~Peg  & T & 2019/06/03 & 540              &  16 \\
       & T & 2019/06/14 & 540              &  14 \\
       & T & 2019/06/26 & 540              &  15 \\
ST~Sgr & T & 2019/04/24 & 2063             &   7 \\
       & T & 2019/05/07 & 1800             &   7 \\
R~Tau  & T & 2019/02/21 & $5\times600$     &  22 \\
\hline
\end{tabular}
\tablefoot{Instrument: T: TIGRE/HEROS; M: Mercator/HERMES}
\end{table}

\section{Analysis and results}\label{sec:analysis}

\subsection{Tc classification}\label{sec:Tc_class}

The reduced TIGRE/HEROS spectra were first shifted to the rest frame by measuring the heliocentric radial velocity (RV) from them. The wavelength range 4200--4330\,\AA\ was used in most cases. Only the spectrum of ST~Sgr had too low an S/N in this range to be useful; therefore, the range 4880--4980\,\AA\ was chosen. We applied a cross-correlation technique with a synthetic spectrum from \citet{2013A&A...553A...6H} as a template. A model with $T_{\rm eff}=3000$\,K and $\log(g) = 0.0$ at solar metallicity was chosen because it is expected to represent the sample stars adequately \citep{2011A&A...531A..88U,2012MNRAS.427..343M}. The synthetic spectrum based on that model atmosphere was convolved with a Gaussian to reduce the spectral resolution to that of the respective observation. The RV correction for the Mercator/HERMES spectra was determined and applied using the standard method in the reduction pipeline.

\begin{figure*}
\centering
\includegraphics[width=\linewidth,bb=54 38 460 326]{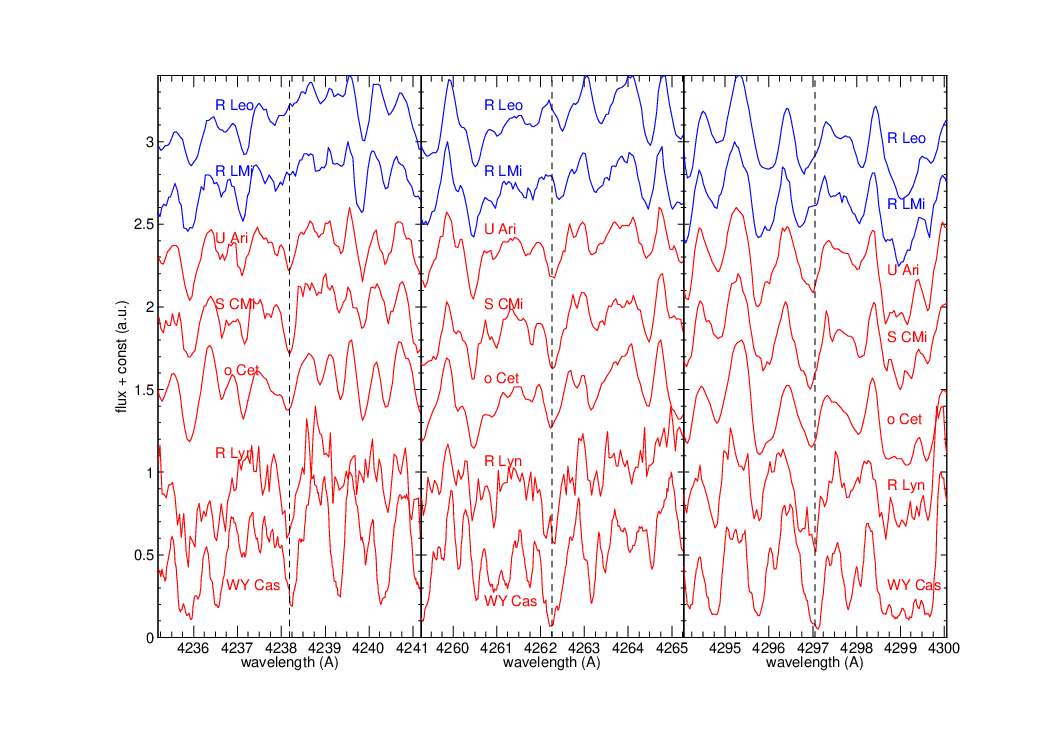}
\caption{Illustration of the Tc lines in a selection of stars. {\it Left panel:} 4238.191\,\AA\ \ion{Tc}{i} line. {\it Middle panel:} 4262.270\,\AA\, \ion{Tc}{i} line. {\it Right panel:} 4297.058\,\AA\ \ion{Tc}{i} line. A dashed vertical line at each panel centre marks the laboratory wavelength of the Tc transitions. The stars from top to bottom are R~Leo, R~LMi, U~Ari, S~CMi, $o$~Cet, R~Lyn, and WY~Cas. Spectra of stars classified as Tc-poor are plotted as blue graphs, and those classified as Tc-rich as red graphs. For clarity, the spectra are shifted by multiples of 0.4 arbitrary flux units along the y-axis.}
\label{Fig:Tc-lines-examples}
\end{figure*}

Before further analysis, individual exposures taken on the same night were co-added where necessary. The individual spectra of R~Peg, observed on different nights, also have relatively low S/N. We shifted them to the rest frame and co-added them to increase the final S/N. We searched the final spectra for the resonance \ion{Tc}{i} lines at 4238.191, 4262.270, and 4297.058\,\AA\ \citep{1968NISTJ..72A.559B} by visually inspecting and comparing them to those of stars known to have (e.g.\ $o$~Cet) or have no Tc (e.g.\ g~Her) and for which TIGRE spectra are also available. We also applied the flux-ratio method introduced by \citet{2007A&A...463..251U}. In this method, the mean flux in a narrow region around a nearby quasi-continuum point is divided by the mean flux in a few wavelength points around the laboratory wavelengths of the Tc lines. Tc-rich stars are identified by having enhanced ratios in all three lines, clearly separating them from Tc-poor ones. We also inspected the subordinate lines in the higher-resolution Mercator/HERMES spectra (3984.967, 4031.626, 4049.108, 4088.707, and 4095.668\,\AA). These are not so clearly detected but essentially confirm the results obtained from the resonance lines. Figure~\ref{Fig:Tc-lines-examples} shows the spectra around the Tc resonance lines of a selection of stars. The stars classified as Tc-rich (red graphs) have an absorption centred around the laboratory wavelengths of Tc (dashed vertical lines), which is lacking in the Tc-poor ones (blue graphs). R~Lyn and WY~Cas have extremely strong Tc lines. R~Leo has already been classified in the literature: \citet{1987AJ.....94..981L} report it as 
'Tc doubtful', and \citet{1999A&A...351..533L} report it as 'Tc no', which we confirm here.

U~Her is interesting because it is classified as Tc-poor, but has a significant Li line absorption. It has a pulsation period of 407\,d and SIMBAD\footnote{\url{https://simbad.u-strasbg.fr/simbad/}} classifies it as an OH/IR star. Adopting its Gaia DR3 \citep{2023A&A...674A...1G} parallax of $\varpi=2.3568\pm0.0770$\,mas and applying a fit to its spectral energy distribution with the PySSED routine \citep{2024RASTI...3...89M}, we determine a luminosity of $L_{\star}\approx18\,000\,L_{\sun}$. All of this indicates that U~Her could be an intermediate-mass star undergoing hot bottom burning similar to those presented in \citet{2013A&A...555L...3G}.

The results of the RV determination and the search for Tc are reported in Table~\ref{tab:results}, which is an updated version of Table~A.2 of paper~II for the objects discussed here. Average values for $\dot{M}_{\rm gas}$ and $v_{\infty}$ have been adopted if observation of more than one CO line is reported in the literature.

\begin{table*}
\caption{Observational results and parameters of the new sample stars.}
\label{tab:results}
\centering
\begin{tabular}{lcrcrrrrrr}
\hline\hline
Name   & Tc & RV$_{\rm helio}$ & $P$ & SpType & $K_{\rm S}$ & $W4$ & $\dot{M}_{\rm gas}$     & $v_{\infty}$ & Ref. \\
       &    & (km\,s$^{-1}$)   & (d) &        & mag         & mag  & $M_{\sun}{\rm yr}^{-1}$ & km\,s$^{-1}$ &      \\
\hline
(1)    & (2) & (3)    & (4)   & (5) & (6)      & (7)      & (8)                & (9)  & (10) \\
\hline
T Ari  & 0  &  $+6.8$ & 313 &  M  &   0.172  & $-1.506$ & $2.3\times10^{-8}$ &  3.0 & d \\
U Ari  & 1  & $-44.6$ & 372 &  M  &   1.435  & $-1.055$ & $6.5\times10^{-8}$ &  4.0 & d \\
R Aur  & 1  &      -- & 457 &  M  & $-0.690$ & $-2.814$ & $1.1\times10^{-6}$ & 10.2 & d \\
U Aur  & 0  &      -- & 409 &  M  &   1.021  & $-2.005$ & $4.4\times10^{-7}$ &  9.5 & d \\
R Cas  & 0  &      -- & 431 &  M  & $-1.404$ & $-4.825$ & $2.3\times10^{-6}$ & 13.0 & c \\
WY Cas & 1  &      -- & 481 &  S  &   1.864  & $-1.740$ & $1.1\times10^{-6}$ & 13.5 & b \\
S CMi  & 1  & $+74.6$ & 334 &  M  &   0.520  & $-1.498$ & $4.5\times10^{-8}$ &  3.5 & a \\
W Cnc  & 0  & $+45.4$ & 394 &  M  &   1.049  & $-1.559$ & $3.1\times10^{-8}$ &  4.2 & d \\
U Her  & 0  &      -- & 407 &  M  & $-0.342$ & $-3.076$ & $3.7\times10^{-7}$ &  9.5 & d \\
X Hya  & 0  & $+42.3$ & 301 &  M  &   0.675  & $-1.435$ & $3.0\times10^{-8}$ &  5.9 & d \\
R Leo  & 0  & $+11.0$ & 305 &  M  & $-2.549$ & $-4.773$ & $2.6\times10^{-7}$ &  7.8 & a \\
R LMi  & 0  & $+10.0$ & 373 &  M  & $-0.224$ & $-3.124$ & $2.5\times10^{-7}$ &  8.1 & a \\
R Lyn  & 1  & $+25.5$ & 379 &  S  &   2.140  &   0.559  & $3.3\times10^{-7}$ &  7.5 & b \\
X Oph  & 0  & $-74.4$ & 334 &  M  & $-0.914$ & $-2.933$ & $1.8\times10^{-7}$ &  5.7 & e \\
R Peg  & 0  & $+23.8$ & 378 &  M  &   0.482  & $-2.067$ & $3.7\times10^{-7}$ &  7.6 & e \\
ST Sgr &(1) & $+47.9$ & 394 &  S  &   1.020  & $-0.772$ & $2.0\times10^{-7}$ &  6.0 & b \\
R Tau  & 0  & $+34.7$ & 323 &  M  &   0.758  & $-1.627$ & $3.1\times10^{-8}$ &  5.3 & f \\
SV Cas & 0  &      -- & 444 &  M  &   0.129  & $-2.380$ & $2.5\times10^{-7}$ &  8.6 & g \\
SY Mon & 0  &      -- & 425 &  M  &   0.848  & $-1.869$ & $3.0\times10^{-6}$ & 11.5 & g \\
\hline
\end{tabular}
\tablefoot{Meaning of the columns: (1): variable name; (2): Tc content (0 = Tc absent, 1 = Tc present); (3): heliocentric radial velocity; (4): pulsation period; (5): spectral type; (6): mean $K_{\rm S}$-band magnitude; (7): WISE~4 magnitude; (8): gas MLR; (9): terminal expansion velocity; (10): reference for the gas MLR and $v_{\infty}$. References: (a): \citet{1985ApJ...292..640K}; (b): \citet{1986ApJ...311..731K}; (c): \citet{1988A&A...196L...1O}; (d): \citet{1995ApJ...445..872Y}; (e): \citet{1998ApJS..117..209K}; (f): \citet{2002A&A...391..577S}; (g): \citet{2022MNRAS.512.1091S}.}
\end{table*}

We also searched the data of the NESS project \citep{2022MNRAS.512.1091S} for new CO observations of Miras with existing Tc observations, finding CO detections of SV~Cas and SY~Mon. Both are reported to be Tc-poor by \citet{2013A&A...555L...3G} and \citet{2003A&A...411..533L}, respectively.

The observations presented in this paper were done to increase the sample of Miras with information about their gas MLR and Tc content (i.e.\ 3DUP activity). We now have 137 Miras with information about their gas MLR and 173 Miras with information about their Tc content. For a subsample of 56 Miras, we know both; 19 are added by this paper (Table~\ref{tab:results}). 

\subsection{Zirconium monoxide}\label{sec:ZrO}

\begin{figure}
\centering
\includegraphics[width=\linewidth,bb=26 10 486 346]{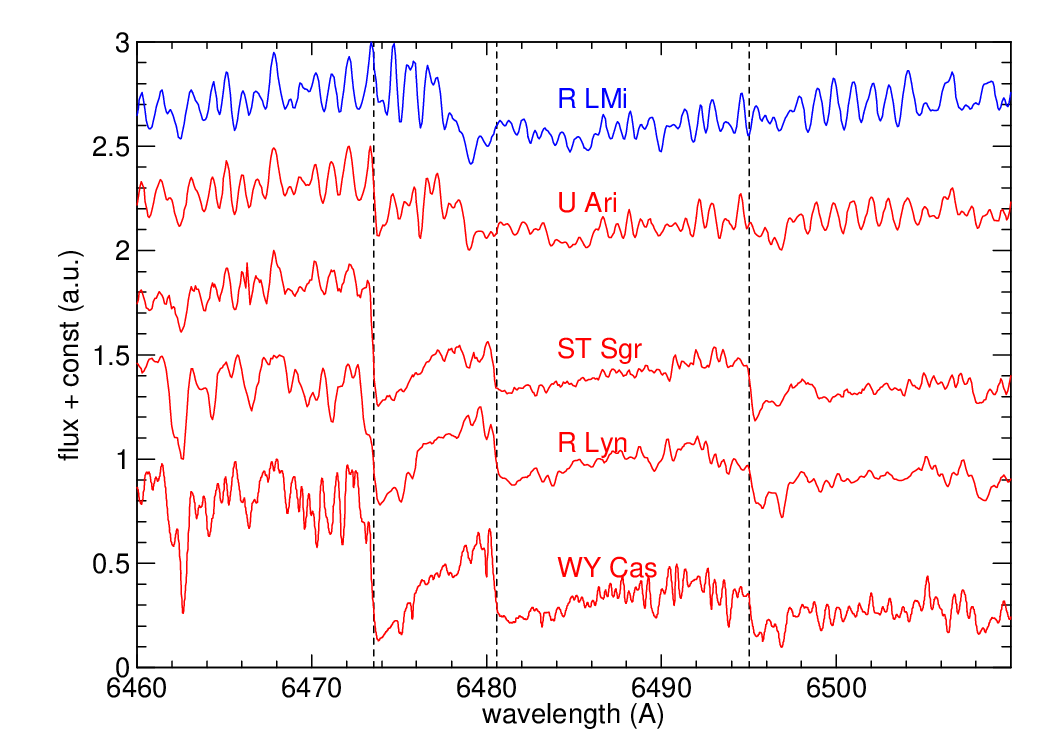}
\caption{ZrO band heads of five selected sample stars. The stars from top to bottom are R~LMi, U~Ari, ST~Sgr, R~Lyn, and WY~Cas. Spectra of Tc-poor stars are plotted as blue and Tc-rich stars as red graphs. For clarity, the spectra are shifted by multiples of 0.5 arbitrary flux units along the y-axis. The wavelengths of ZrO band heads, taken from \citet{1947ApJ...106...28D}, are marked by dashed vertical lines. From left to right: ZrO $\gamma(0,0)$ Rc (6473.54\,\AA), $\gamma(0,0)$ Xc (6480.57\,\AA), and $\gamma(1,1)$ Rc head (6495.00\,\AA). ST~Sgr is discussed in more detail in Sect.~\ref{sec:ZrO}.}
\label{Fig:ZrO}
\end{figure}

Zirconium (Zr) is also an important indicator of $s$-process enrichment, and it is most readily detectable in stellar spectra through the bands of the ZrO molecule \citep[e.g.][]{2017A&A...601A..10V}. In S stars, these bands dominate the spectral appearance and thus define this spectral type. Figure~\ref{Fig:ZrO} compares the ZrO bands around $\sim6500$\,\AA\ of five sample stars: one Tc-poor and four Tc-rich ones. The Tc-rich star with the weakest ZrO bands shown, U~Ari, is classified as a pure M-type star \citep{2014yCat....1.2023S}. Nevertheless, our spectra demonstrate that, despite being a pure M-type star, it does have weak but significant ZrO absorption, as is evident from the drop in flux just redward of the $\gamma(0,0)$ Rc band head (6473.54\,\AA) in Fig.~\ref{Fig:ZrO}. This confirms earlier findings by \citet{2011A&A...531A..88U} that Tc-rich pure M-type stars do have detectable ZrO bands. On the other hand, none of the stars in our sample classified as Tc-poor has ZrO absorption, confirming that they have not undergone $s$-process enrichment.

However, Zr is a radioactively stable element, so it could also have been enriched by mass transfer from a former AGB binary companion instead of internal nucleosynthesis and mixing. Unfortunately, both spectra of ST~Sgr have an S/N that is too low in the wavelength range of the Tc lines, making it impossible to detect those lines. The red part of the spectra has a high S/N, so we can confirm the presence of strong ZrO bands in ST~Sgr (Fig.~\ref{Fig:ZrO}), but they could have their origin in a previous mass-transfer event. Therefore, ST~Sgr could be an extrinsic S star, not an intrinsic one. However, good reasons exist to believe it is an intrinsic S star.

First, intrinsic S stars have been shown to be photometrically much more variable than extrinsic S stars \citep[e.g.][]{1972ApJ...177..489E}. In accordance with this, ST~Sgr is a Mira variable with an enormous amplitude of $\sim8\fm0$ in the $V$-band. Second,  depending on stellar mass and metallicity, the onset of 3DUP events occurs at different luminosities on the AGB, depending on mass, but roughly at $L_{\star}\approx2000\,L_{\sun}$. Observations show that intrinsic S stars are typically found at luminosities higher than this limit. In contrast, extrinsic S-type stars, except maybe for very high masses, are found below this luminosity \citep{2021A&A...650A.118S}. Applying the PySSED routine for a spectral energy distribution (SED) fit and adopting ST~Sgr's Gaia DR3 parallax of $\varpi=1.8435\pm0.1429$\,mas, we derive its luminosity to be $L_{\star}\approx6\,500\,L_{\sun}$, putting it in the luminosity range of intrinsic S stars. Finally, based on its IRAS colours, \citet{2019AJ....158...22C} also classify ST~Sgr as an intrinsic S star. Therefore, we may quite safely assume that ST~Sgr is an intrinsic S star that acquired its $s$-element overabundance by internal nucleosynthesis and mixing, not due to mass transfer from a former AGB star companion. For the reasons above, we include ST~Sgr in the analysis as a Tc-rich star, but put the classification in parentheses in Table~\ref{tab:results}.

\subsection{Third dredge-up and mass-loss properties}\label{sec:3DUP_ML}

\begin{figure}
\centering
\includegraphics[width=\linewidth,bb=40 10 486 342]{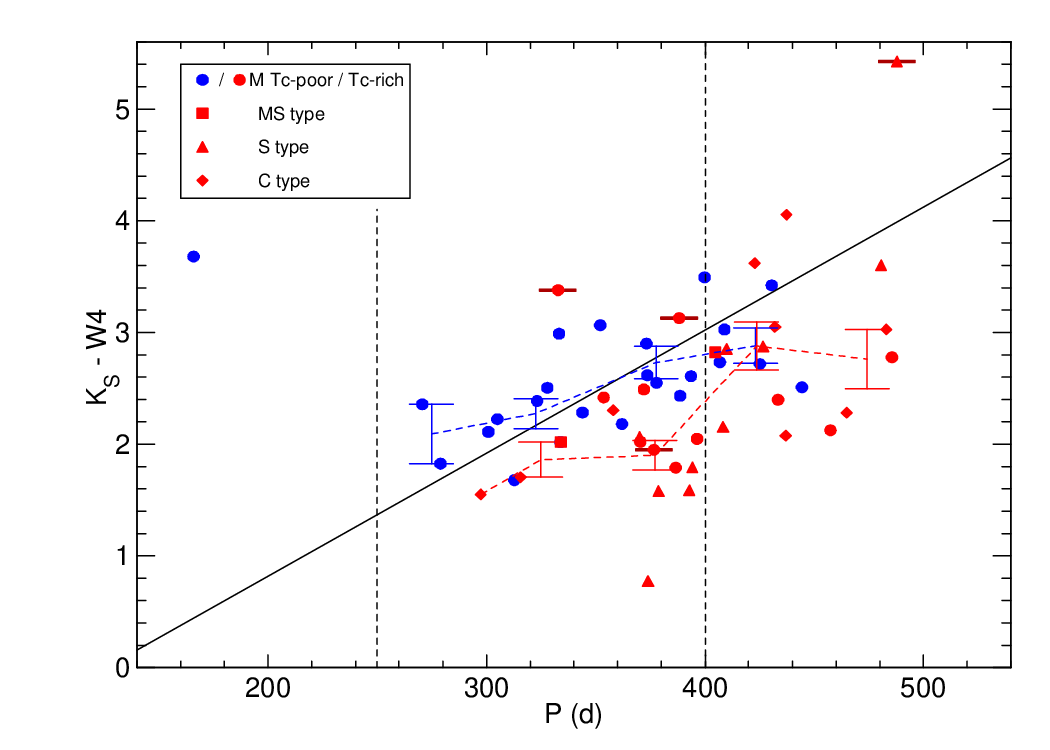}
\caption{$K_{\rm S}-W4$ vs.\ $P$ diagram of the stars with Tc and ML observations.
Tc-poor stars are shown in blue, and Tc-rich ones in red. Different symbols distinguish spectral subtypes; see the legend. The dashed vertical lines mark the period range 250 -- 400\,d, in which the dichotomy between Tc-poor and Tc-rich Miras was found to be most apparent in paper~II. The solid line is optimised to best separate Tc-poor and Tc-rich Miras; see also paper~II. The run of the mean $K-W4$ colour in 50\,d period bins is represented by the blue and red dashed lines with error bar symbols for Tc-poor and Tc-rich Miras, respectively. The size of the vertical bar is scaled to the standard deviation of the mean. The binaries, marked by horizontal dashes, have been excluded from calculating the means; sorted by increasing period, these are: $o$~Cet, R~Hya, R~Aqr, and W~Aql.}
\label{Fig:K-W4_vs_P}
\end{figure}

We now combine the information about the 3DUP activity of our sample stars with their ML properties. We can expect to find a difference between Tc-poor and Tc-rich Miras in $\dot{M}_{\rm gas}$ only if the sample reproduces the dichotomy observed in $K-W4$ and other near-to-mid IR colours. Therefore, we first inspect the $K-W4$ vs.\ $P$ diagram in Fig.~\ref{Fig:K-W4_vs_P}. The solid line in that figure is the one that was optimised in paper~II to separate Tc-poor and Tc-rich Miras best. Indeed, the Tc-poor Miras (blue symbols) accumulate above, or at most slightly below, this line. In contrast, on average, the Tc-rich stars of various spectral types (red symbols) have considerably lower $K-W4$ colours. This impression is confirmed by looking at the mean colour in 50\,d period bins, as shown by the red and blue dashed lines and error bar symbols in Fig.~\ref{Fig:K-W4_vs_P}: The mean $K-W4$ of Tc-poor Miras is significantly redder than that of Tc-rich ones in the same period bin. For example, in the bin $300-350$\,d, $\langle K-W4\rangle=2.27\pm0.14$ for the Tc-poor stars and $\langle K-W4\rangle=1.86\pm0.16$ for the Tc-rich stars, and in the $350-400$\,d bin, $\langle K-W4\rangle=2.73\pm0.14$ for the Tc-poor stars and $\langle K-W4\rangle=1.90\pm0.13$ for the Tc-rich stars (standard deviation of the mean).

Three of the five Tc-rich Miras above the best separating line are well-known binary stars: $o$~Cet, R~Aqr, and W~Aql. The fourth (suspected) binary, R~Hya \citep{2021A&A...651A..82H}, has an inconspicuous $K-W4$ colour. All four binaries have been excluded from calculating the means. Thus, although not as clear-cut as found in the full sample of paper~II, the sample stars with Tc and ML observations confirm the dichotomy in $K-W4$ between the two groups of Miras.

\subsubsection{The gas mass-loss rate}\label{sec:gas-MLR}

From the difference in mean $K-W4$ colour estimated in the previous paragraph, we can estimate the difference in the gas MLR that we can expect to find for a fixed dust-to-gas ratio. For example, applying the relation of \citet{2020NatAs...4.1102M} for M-type stars to the $\left< K-W4\right>$ colours of Tc-poor and Tc-rich Miras in the period range 350 -- 400\,d (see above), this translates to a mean $\log(\dot{M}_{\rm gas})=-7.92$ for the Tc-rich Miras and $\log(\dot{M}_{\rm gas})=-6.85$ for the Tc-poor ones. However, this relation is very steep in the $K-W4$ range of interest here.
\citet{2018MNRAS.481.4984M}, on the other hand, fit the linear relation $\log(\dot{M}_{\rm gas})=0.469\times(K-W4)-7.383$ to their data. This relation gives mean $\log(\dot{M}_{\rm gas})=-6.49$ and $-6.10$, respectively. \citet{2008A&A...487..645R} estimate an uncertainty in the ML determination by a factor of 3. Therefore, uncovering such differences in the $\log(\dot{M}_{\rm gas})$ vs.\ $P$ diagram might be difficult. On the other hand, the standard deviation of the mean will decrease by increasing the number of stars in the compared groups.

\begin{figure}
\centering
\includegraphics[width=\linewidth,bb=33 10 486 342]{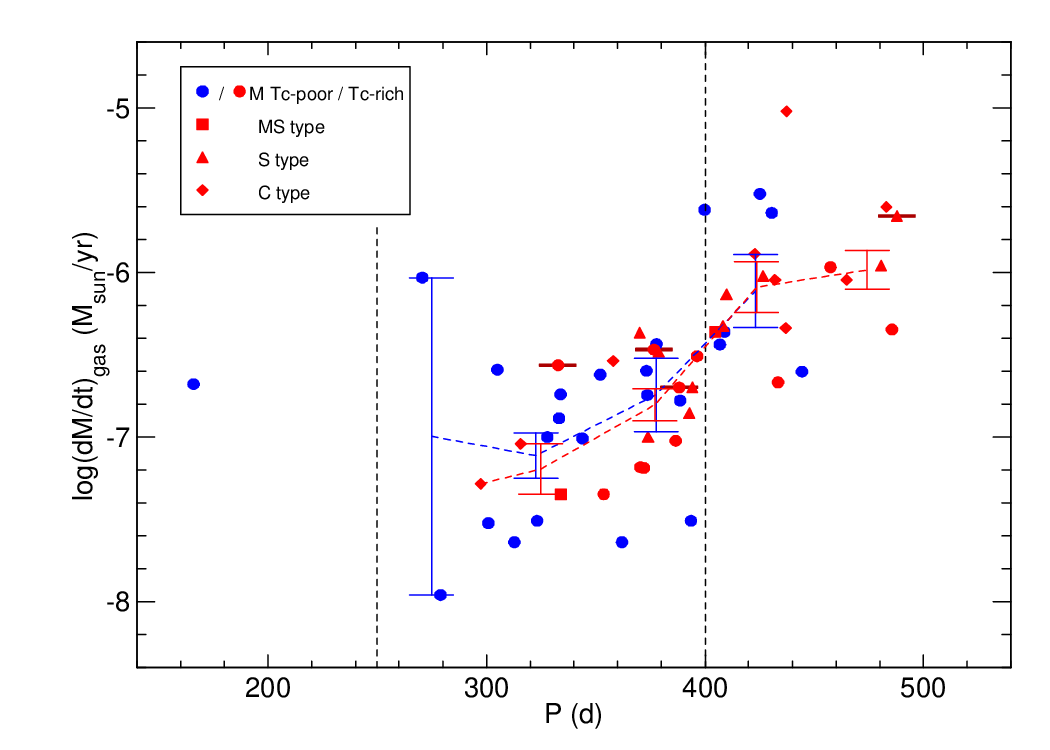}
\caption{$\log(\dot{M}_{\rm gas})$ vs.\ $P$ diagram. Symbols are as in Fig.~\ref{Fig:K-W4_vs_P}. At a given period, the gas MLRs of Tc-poor and Tc-rich stars are indistinguishable; see Sect.~\ref{sec:gas-MLR}.
}
\label{Fig:Mdotgas_vs_P}
\end{figure}

Figure~\ref{Fig:Mdotgas_vs_P} shows the $\log(\dot{M}_{\rm gas})$ vs.\ $P$ diagram of the stars with Tc and ML observations. The Tc-rich objects (red symbols) seem to follow a relatively tight relation of $\log(\dot{M}_{\rm gas})$ as a function of $P$, whereas there is considerable scatter among the Tc-poor stars (blue symbols). Again, we over-plot the run of the mean $\log(\dot{M}_{\rm gas})$ in period bins of 50\,d. Indeed, the mean $\log(\dot{M}_{\rm gas})$ of the Tc-poor Miras is slightly higher than that of the Tc-rich ones in the period range of interest. However, the means differ by less than 0.1\,dex, much less than the standard deviation of the mean (blue and red error bar symbols). Thus, we do not claim this to be a detection: There is no dichotomy between Tc-poor and Tc-rich Miras in $\log(\dot{M}_{\rm gas})$, even with the new data.

\subsubsection{The terminal expansion velocity}\label{sec:v_exp}

Besides the MLR, the terminal expansion velocity $v_{\infty}$ of the stellar wind is another important ML parameter. We therefore also inspect the $v_{\infty}$ vs.\ $P$ diagram. Here, we have further restricted the selection of the stars by excluding the carbon stars. This selection is motivated by the fact that the kinetic coupling between dust and gas may differ between O-rich and C-rich stars, and they can have very different drift velocities (velocity difference between dust and gas). However, we want to inspect differences in (gas) expansion velocity between Tc-poor and Tc-rich stars, not between the different chemical regimes. Since the dust emission does not have such sharp spectral features as atoms and molecules do, the expansion velocity of the dust cannot be measured directly from observations, and the drift velocity is unknown. According to \citet{2021Univ....7..113M}, the drift velocity $v^{2}_{\rm D}$ can be approximated by
\begin{equation}
    v^{2}_{\rm D}\approx\frac{\left<Q_{\rm ext}\right>L_{\star}}{\dot{M}c}u,
\end{equation}
where $\left<Q_{\rm ext}\right>$ is the relative extinction cross-section averaged over the entire spectrum, $L_{\star}$ is the bolometric luminosity of the star, $\dot{M}$ is the mass-loss rate, $c$ the speed of light, and $u$ is the gas velocity. $\left<Q_{\rm ext}\right>$ depends on the material properties and the size distribution of the dust grains, which are not known very well. For small grains of amorphous carbon around C stars, it is reasonable to accept $\left<Q_{\rm ext}\right>=1$ \citep{2003A&A...400..981A}. In the oxygen-rich case, the chemical composition, stoichiometry, and grain size can vary from star to star, so $\left<Q_{\rm ext}\right>$ is not well-constrained. Nevertheless, it is reasonable to assume that $\left<Q_{\rm ext}\right>$ is smaller in the oxygen-rich regime than in the carbon-rich one, changing as a star evolves from one regime to the other. On the other hand, the effect of taking into account drift in the wind modelling has stronger effects in O-rich than in C-rich stars \citep{2023A&A...677A..27S}. The O-rich stars (including S stars) contain Tc-poor and Tc-rich objects. Thus we can compare these groups in the analysis. 

\begin{figure}
\centering
\includegraphics[width=\linewidth,bb=26 10 486 342]{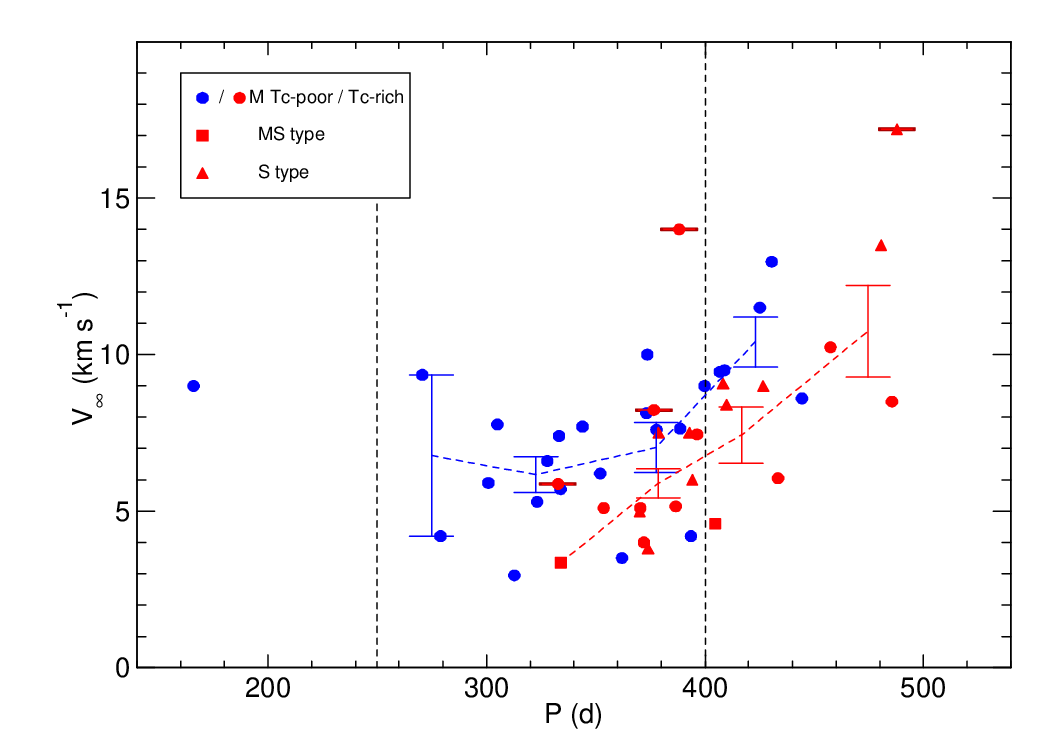}
\caption{$v_{\infty}$ vs.\ $P$ diagram of the O-rich stars with Tc and ML observations. Symbols are as in Fig.~\ref{Fig:K-W4_vs_P}. At a given pulsation period, the Tc-poor Miras appear to have a higher expansion velocity than the Tc-rich ones; see Sect.~\ref{sec:v_exp}.}
\label{Fig:vexp_vs_P}
\end{figure}

Figure~\ref{Fig:vexp_vs_P} presents the $v_{\infty}$ vs.\ $P$ diagram. Again, the average $v_{\infty}$ in 50\,d bins is plotted with the standard deviation of the mean. It seems that the Tc-poor Miras (blue symbols) have a higher terminal expansion velocity than the Tc-rich ones (red symbols) over a range of periods, reminiscent of the findings in the $K-W4$ colour. In this restricted sample, there is only one Tc-rich Mira between 300 and 350\,d of period; it does have a lower expansion velocity than all but one of the Tc-poor Miras in the same period interval. In the interval 350 -- 400\,d, the average $v_{\infty}$ of the Tc-poor Miras is also higher than that of the Tc-rich ones, but the error bars overlap slightly. Finally, all but one Tc-poor stars have a higher $v_{\infty}$ than the Tc-rich stars in the interval 400 -- 450\,d.

The known binary stars also appear to have enhanced $v_{\infty}$. In particular, R~Aqr and W~Aql have the highest expansion velocities of all stars in Fig.~\ref{Fig:vexp_vs_P}. This fact appears to confirm hydro-dynamical simulations such as those of \citet{2021A&A...653A..25M} or \citet{2022MNRAS.513.4405A}, who showed that close (sub)stellar companions can increase the terminal wind velocity or MLR w.r.t.\ a single AGB star by a gravity assist on the wind particles. We note, however, that these models do have some limitations. Most importantly, they assume that dust and gas are mechanically fully coupled (one-fluid approach). 

A few Tc-poor Miras with very low expansion velocities ($v_{\infty}<5\,{\rm km}\,{\rm s}^{-1}$) are included in the calculation of the means, namely T~Ari, R~Cnc, and W~Cnc. The existence of a small population of stars with such low expansion velocities is surprising; see, e.g.\ \citet{1993A&AS...99..291L} and \citet{1999A&AS..138..299K} for a discussion of this group. \citet{2000A&A...361..641W} model low-expansion-velocity winds and find that they are consistent with their B-type models; that is, the mass loss is driven by the pulsations rather than the radiation pressure on dust grains. Indeed, the three mentioned stars lie on the blue edge of the Tc-poor sequence in the $K-W4$ vs.\ $P$ diagram. However, they are not bluer than Tc-rich Miras at similar periods. Finally, note that some low-expansion-velocity objects, including T~Ari, have been found to show SiO maser features with velocity distributions much broader than the CO profiles, which is interpreted as the result of two (or more) successive winds with very different MLRs and velocities \citep{2003A&A...409..715W}. 

Figure~\ref{Fig:vexp_vs_P} suggests a dichotomy between Tc-poor and Tc-rich Miras in $v_{\infty}$, just as in the $K-W4$ colour. This dichotomy is in the same sense in both quantities, namely that, at a given period, Tc-poor Miras have higher $v_{\infty}$ and $K-W4$ than their Tc-rich siblings. It also indicates that Tc-poor Miras have more circumstellar dust (or a higher dust emissivity; see below) and a more efficient momentum transfer from the dust grains to the molecular gas than Tc-rich Miras. However, this needs to be confirmed by a larger sample and improved CO data before we can regard this as firmly established. 


\subsection{Infrared colours of Miras and their modelling}\label{sec:IR-colours}

\subsubsection{WISE colours versus\ period diagram and DARWIN models}\label{sec:W1-W4_vs_P}

The WISE All-Sky Data Release \citep{2010AJ....140.1868W} provides a very valuable database for analysing the MIR dust emission of Miras. With its four bands, WISE covers a very broad wavelength range that includes bands in which the spectrum of a typical optical Mira star is dominated by the light coming from the photosphere ($W1$, $\lambda_{\rm e}=3.4\,\mu{\rm m}$) and bands in which the circumstellar dust emission can dominate ($W4$, $\lambda_{\rm e}=22.2\,\mu{\rm m}$). Thus, similar to $K-W4$, the $W1-W4$ colour is also an indicator of the dust MLR of a Mira. Additional advantages are that the observations are taken quasi-simultaneously in the various bands, thus minimising the impact of stellar variability on the colours, and that the $W1$ band is marginally less subject to pulsation-induced variability than the $K$-band. Unfortunately, nearby Miras are so bright that most are saturated in $W1$. According to \citet{2012wise.rept....1C}, "WISE profile-fitting photometry reliably extracts measurements of saturated sources using the non-saturated wings of their profiles up to a brightness of approximately 2\fm0 in $W1$". Restricting the sample of stars with Tc observations to distant, faint Miras with $W1$ photometry reliably determined in this way, we obtain 36 Tc-poor and 29 Tc-rich Miras, the brightest of which as $W1=2\fm317$. 

\begin{figure}
\centering
\includegraphics[width=\linewidth,bb=40 10 486 342]{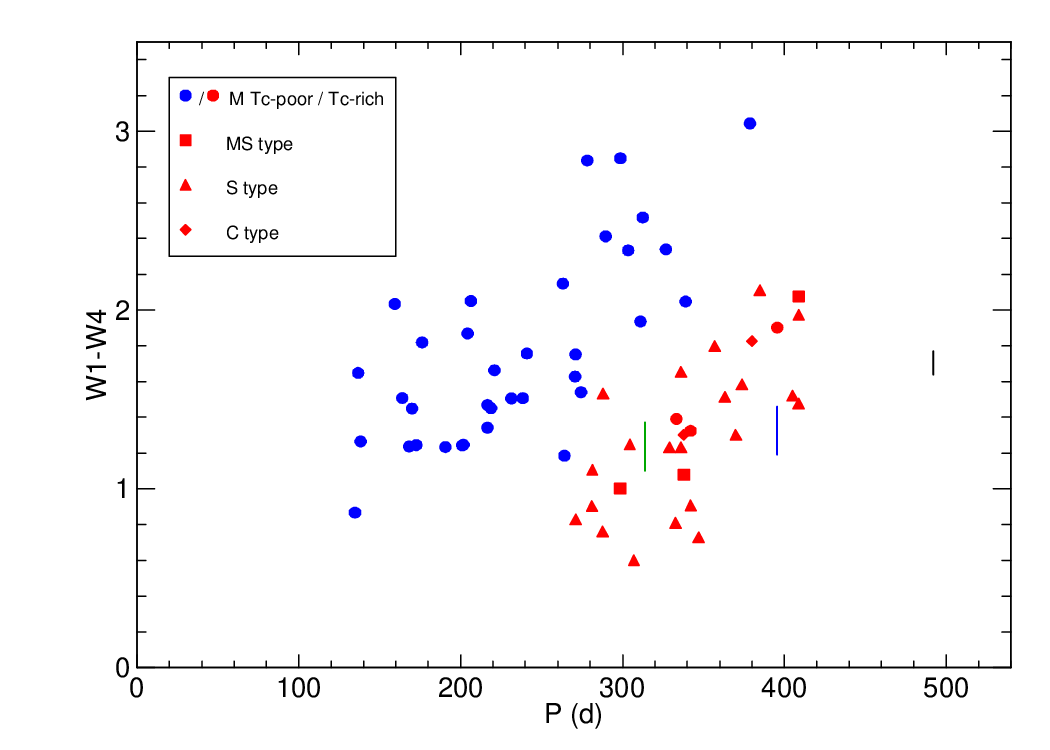}
\caption{$W1-W4$ vs.\ $P$ diagram of the Miras with Tc observation and $W1$ photometry in the WISE All-Sky Data Release. Symbols are as in Fig.~\ref{Fig:Mdotgas_vs_P}. For comparison, the vertical lines show the maximum colour variation of three DARWIN dynamical wind models with Fe-rich dust: An114u4 (green), Bn114u4 (blue), and M2n315u6 (black). The model parameters are summarised in Table~\ref{tab:DARWIN}.}
\label{Fig:W1-W4_vs_P}
\end{figure}

The $W1-W4$ vs.\ $P$ diagram of this faint Mira sample is displayed in Fig.~\ref{Fig:W1-W4_vs_P}. As is evident from this figure, Tc-poor and Tc-rich Miras are clearly separated in this diagram. In the period interval where the two groups overlap ($\sim270-380$\,d), Tc-poor Miras have a higher colour index than Tc-rich ones, such that both groups form separate sequences for which $W1-W4$ increases with pulsation period $P$. This dichotomy strongly suggests that Miras that have not (yet) experienced 3DUP either have a much higher dust column density (indicative of a higher dust MLR), or a higher dust emissivity.

The separation in this $W1-W4$ vs.\ $P$ diagram is even cleaner than in the $K-W4$ vs.\ $P$ diagram (see paper~II). This fact suggests that stellar variability is still a significant source of scatter in the latter diagram, though other effects could be at work in broadening the sequences. We note here that the data used to construct this intriguing diagram stems from very different sources: The IR photometry comes from the WISE space observatory, the pulsation periods are mainly based on visual AAVSO light curves, and the information on the Tc content is based on spectral observations, taken from various literature sources and our own work (we refer to paper~II for references). Thus, we rule out any possibility that the two sequences are a chance alignment.

The vertical lines over-plotted in Fig.~\ref{Fig:W1-W4_vs_P} represent the colour variation of dynamical atmosphere and wind models of M-type AGB stars calculated with the DARWIN code \citep{2016A&A...594A.108H,2022A&A...657A.109H}. The radial structures of the models are obtained by solving the equations of radiation-hydrodynamics together with a time-dependent description of dust grain growth. Starting from dust-free hydrostatic structures that are defined by the fundamental stellar parameters, the effects of stellar pulsation are introduced by varying the luminosity and gas velocity at the inner boundary, located just below the photosphere. The models used here are taken from \citet{2022A&A...657A.109H}, and they include self-regulating Fe-enrichment of the silicate grains. The model parameters and the resulting wind properties are summarised in Table~\ref{tab:DARWIN}. We note that the predicted combinations of MLRs and wind velocities, as well as the visual and NIR colours, are in good agreement with observations \citep[see Figs.~5 and 6 in ][]{2022A&A...657A.109H}. To obtain synthetic spectra and photometric fluxes, snapshots of the DARWIN models are post-processed with the COMA radiative transfer code \citep{2016MNRAS.457.3611A}.

The models with Fe-enriched silicate dust reach higher $W1-W4$ colours than those with completely transparent, Fe-free Mg$_2$SiO$_4$ dust that is near-transparent at NIR wavelengths. Indeed, they reasonably well reproduce the colours of the bulk of the sample. However, they do not reach the reddest $W1-W4$ colours, particularly of the Tc-poor Miras with the longest periods. Even the model M2n315u6, with a pulsation period much longer than the observed sample stars, only has a slightly redder $W1-W4$ index than the shorter-period models. We note here that the models with Fe-bearing silicate grains are currently only available for a very limited range of stellar parameters.


\begin{table*}
\caption{Parameters and resulting wind properties of the DARWIN models.}
\label{tab:DARWIN}
\centering
\begin{tabular}{lcccccccc}
\hline\hline
Name     & $M$        & $L$       & $T_{\rm eff}$ & $P$ & $\Delta u$   & $n_{\rm d}/n_{\rm H}$ & $\dot{M}$               & $v_{\infty}$ \\
         & $M_{\sun}$ & $L_{\sun}$& K             & d   & km\,s$^{-1}$ &                       & $M_{\sun}{\rm yr}^{-1}$ & km\,s$^{-1}$ \\
\hline
(1)      & (2)        &  (3)      & (4)           & (5) & (6)                   & (7)                     & (8)          & (9)          \\
\hline
An114u4  & 1.0        & 5000      & 2800          & 310 & 4            & $10^{-14}$            & $5\times10^{-7}$        &  9           \\
Bn114u4  & 1.0        & 7000      & 2700          & 390 & 4            & $10^{-14}$            & $8\times10^{-7}$        & 10           \\
M2n315u6 & 1.5        & 7000      & 2600          & 490 & 6            & $3\times10^{-15}$     & $2\times10^{-6}$        & 11           \\
\hline
\end{tabular}
\tablefoot{Meaning of the columns: (1): model name; (2): mass; (3): luminosity; (4): effective temperature; (5): pulsation period; (6): piston velocity; (7): seed particle abundance; (8): mass-loss rate; (9): outflow velocity. Mass-loss rate and outflow velocity are results, not input parameters.}
\end{table*}

A closer inspection of the model spectra regarding the relatively modest $W1-W4$ colours revealed that the silicate dust feature matches in strength observed ones in ISO spectra \citep[e.g.\ R~Aql,][]{2003ApJS..147..379S}. However, the flux in the $W1$ filter is significantly affected by photospheric absorption by a water band centred on $2.7$\,$\mu{\rm m}$. At the phases when the model produces strong dust features, the water absorption is the weakest, diminishing the $W1-W4$ colour index. The water absorption is always weaker in the DARWIN models than in observed ISO spectra. Reproducing the strength of this H$_2$O band with atmospheric models is a long-standing difficulty \citep{2002A&A...395..915A}. Some works have invoked a stationary molecular layer 'molsphere') just above the stellar surface to solve this problem \citep[e.g.][]{1997A&A...320L...1T} but this is applicable only to hydrostatic models, not to hydrodynamic models. Furthermore, the synthetic dust features appear narrower at model phases with strong emission. The models may be missing some molecular or dust emission longward of $10$\,$\mu{\rm m}$. The WISE photometric bands are vital tests for AGB wind models, but we must defer a more detailed analysis to future work.

We also observe that both the Tc-poor and Tc-rich Mira sequences have a larger spread in $W1-W4$ than the range of colours covered by the models during the simulated pulsation cycles. Either the models underestimate the cyclic colour variation of real Miras, or the width of the sequences includes a star-to-star scatter on top of the stellar variability.

\subsubsection{WISE colours of stationary wind models}\label{sec:DUSTY}

We also made efforts to model the stars' IR colours with stationary wind models that employ the \textsc{dusty} radiative transfer code\footnote{\url{https://github.com/ivezic/dusty}} \citep{ivezic97}. These models were calculated in the following way. 
We computed dust growth of different dust species coupled with the stationary wind as a function of the stellar parameters resulting from the evolutionary tracks computed with the FUNS code\footnote{\url{fruity.oa-abruzzo.inaf.it}} \citep{2006NuPhA.777..311S,cris09,cris11,pier13,cris15}. A similar procedure has already been adopted in the literature \citep{Ventura12, nanni13, nanni14}. We computed the dust composition, outflow velocity, and colours of selected models along a $1.5 M_{\odot}$ initial mass track of solar metallicity. The luminosity, effective temperature, and MLR varied along the tracks in the ranges $4\,100\lesssim L_{\star}/L_{\odot}\lesssim10\,400$, $2\,700\lesssim T_{\rm eff}\lesssim3\,100\,\mathrm{K}$, and $10^{-7}\lesssim \dot{M}/(M_\odot {\rm yr}^{-1}) \lesssim 4\times10^{-6}$, respectively. We calculated three tracks with progressively higher C abundance, having ${\rm C}/{\rm O}=0.54$, 0.74, and 0.95. The highest ${\rm C}/{\rm O}$ ratio is already quite high but should well illustrate the effects. The calculation of the dust formation (see below) takes into account the reduction of free O with increasing ${\rm C}/{\rm O}$. The input spectra for the calculations were interpolated in effective temperature and ${\rm C}/{\rm O}$ between the solar-metallicity ($\log({\rm Fe}/{\rm H})+12=7.56$) input grids of \citet{2016MNRAS.457.3611A}, with BT2 data for water \citep{barber06}. At each of the three ${\rm C}/{\rm O}$ ratios, we selected models at four different MLRs, namely $\approx2\times10^{-7}$, $\approx3\times10^{-7}$, $\approx1\times10^{-6}$, and $\approx1.5\times10^{-6} M_\odot$\,yr$^{-1}$, to demonstrate the effects of MLR and ${\rm C}/{\rm O}$ on the various quantities in the figures of this section.

We followed the growth of corundum (Al$_2$O$_3$), olivine (Mg$_{2x_{\rm ol}}$Fe$_{2(1-x_{\rm ol})}$SiO$_4$), pyroxene (Mg$_{x_{\rm py}}$Fe$_{(1-x_{\rm py})}$SiO$_3$), and metallic iron. Here, $x_{\rm ol, py}={\rm Mg}_{\rm ol, py}/({\rm Mg}_{\rm ol, py}+{\rm Fe}_{\rm ol, py})$ is the fraction of magnesium in the grain over the total magnesium and iron, and metallic iron. We adopted the optical constants from \citet{dorsch95} with $x_{\rm ol}=x_{\rm py}=0.05$. The initial expansion velocity of the envelope ($v_0=0.1$\,km\,s$^{-1}$) and the seed particle abundance for all the dust species considered ($\epsilon_s=10^{-15}$) were adjusted to reproduce the well-known observed relation between the expansion velocity and mass-loss rate; see discussion below.

For olivine and pyroxene, we assumed that the destruction mechanism is free evaporation. In this case, the destruction rate is computed in analogy to Eq.~3 of \citet{kobayashi11} and outlined in \citet{nanni13, nanni14}. For corundum, we assumed the reaction between H$_2$ molecules and the grain surface, known as chemisputtering, to be fully efficient. The sticking coefficients, $\alpha_i$ (the probability for an atom or molecule to stick on the grain surface), are selected as in \citet{Gail99} and \citet{nanni13}, but for olivine and pyroxene, we chose a sticking coefficient of 0.4, which better reproduces the observed expansion velocities. The dust opacities are consistently computed for the typical grain size obtained at the end of each integration.

The outputs of the growth code coupled with a stationary dust-driven model are used as input for the latest version of the \textsc{dusty} code. The approach adopted is analogous to \citet{nanni16} and \citet{nanni19}. In particular, we assumed: a) the dust-free photospheric spectrum as used for the dust-growth calculation; b) the temperature at the inner boundary of the dust shell to be equal to the condensation temperature of olivine; c) the mass-weighted optical properties of the dust mixture; d) the dust density profile; e) the optical depth at 1\,$\mu{\rm m}$ at the end of the integration. The reaction and input parameters selected for the calculation of the dust condensation are summarised in Table~\ref{tab:parameters}.

\begin{table*}\label{Table:properties}
\caption{Reactions, sticking coefficients, and references of optical constants selected for the calculation of dust condensation in the stationary wind models.}
\begin{tabular}{llll}
\hline
Dust species & Overall formation reaction       & $\alpha_i$    & Optical constants\\
\hline
Corundum        & Al$_2$O+2H$_2$O $\rightarrow$ Al$_2$O$_3$(s)+ 2H$_2$ & $1.0$ & \citet{Begemann97} \\
Olivine         & 2x$_{\rm ol}$Mg+2(1-x$_{\rm ol}$)Fe+SiO+ 3 H$_2$O $\rightarrow$ Mg$_{\rm 2x_{\rm ol}}$Fe$_{\rm 2(1-x_{\rm ol})}$SiO$_4$(s)+3H$_2$ & $0.4$ & \citet{dorsch95} \\
Pyroxene        & x$_{\rm py}$Mg+(1-x$_{\rm py}$)Fe+SiO+ 2 H$_2$O $\rightarrow$ Mg$_{x_{\rm py}}$Fe$_{(1-x_{\rm py})}$SiO$_3$(s)+2H$_2$ & $0.4$ & \citet{dorsch95} \\
Iron            & Fe $\rightarrow$ Fe(s) & $1.0$ & \citet{Ordal88} \\
\hline
\end{tabular}
\label{tab:parameters}
\end{table*}

We cannot plot the stationary wind models in Fig.~\ref{Fig:W1-W4_vs_P} because the models do not provide a pulsation period. Therefore, we decided to analyse the $W1-W4$ vs.\ $J-K_{\rm S}$ colour-colour diagram. We aimed at reproducing the reddest observed $W1-W4$ colours of the stars with Tc observation ($W1-W4\approx3$). Before doing so, we gauged the models to match important mass-loss parameters of the stars, particularly the MLR, the expansion velocity ($v_\infty$), and the $V-K_{\rm S}$ colour. This colour index is sensitive to the stellar effective temperature and to the circumstellar absorption in the optical spectral range, that is, the transparency of the dust grains. 

\begin{figure}
\centering
\includegraphics[width=\linewidth,bb=28 12 232 326]{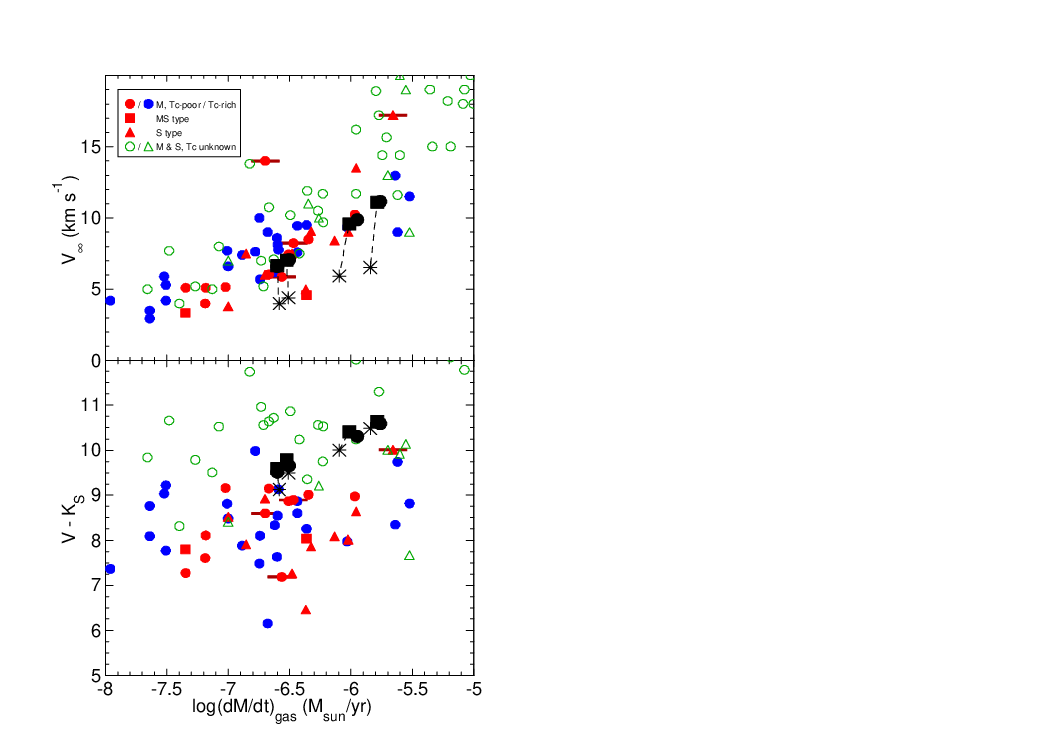}
\caption{$v_\infty$ vs.\ MLR ({\it top panel}) and $V-K_{\rm S}$ vs.\ MLR ({\it bottom panel}). O-rich Miras with Tc observations are shown with the same symbols as in Fig.~\ref{Fig:K-W4_vs_P}. Open green symbols represent Miras without Tc observations. Black symbols represent the based stationary wind models: ${\rm C}/{\rm O}=0.54$: circles; ${\rm C}/{\rm O}=0.74$: squares; ${\rm C}/{\rm O}=0.95$: asterisks. Dashed lines connect models with the same MLR. Both diagrams were used to fine-tune the models to the observed sample stars.}
\label{Fig:DUSTY1}
\end{figure}

The diagrams in Fig.~\ref{Fig:DUSTY1} present the results of our efforts to tune the model parameters by means of the sample stars with ML observations. The top panel shows that the models very well follow the observed trend of increasing outflow velocity with increasing MLR. The tracks do not cover the high MLRs of the stars with the strongest winds but have more moderate rates. While the tracks with ${\rm C}/{\rm O}=0.54$ and 0.74 essentially coincide in expansion velocity, the tracks with ${\rm C}/{\rm O}=0.95$ (asterisks) are clearly shifted to lower values of $v_\infty$. Interestingly, the observed stars indicate the same trend: Up to $\dot{M}=5\times10^{-7} M_\odot {\rm yr}^{-1}$ ($\log(\dot{M})=-6.3$), the Tc-poor stars appear to have higher expansion velocities than the Tc-rich stars, similar to the result of Fig.~\ref{Fig:vexp_vs_P}. It is not possible to decide if the trend continues to higher MLRs due to too few stars. We also note that the two binaries R~Aqr and W~Aql again clearly deviate from the overall trend at much higher $v_\infty$.

The lower panel of Fig.~\ref{Fig:DUSTY1} displays the time-averaged $V-K_{\rm S}$ colour as a function of MLR. Miras with and without Tc differ very little in $V-K_{\rm S}$, indicating that differences in $T_{\rm eff}$ and dust transparency are small. Stars without Tc observations (green symbols) tend to have redder $V-K_{\rm S}$ colours; some are even off-scale. Both panels indicate that the Tc observations are biased to less reddened stars with less extreme stellar winds, as may be expected from the fact that sufficient flux in the blue spectral range is required to search for the Tc lines. The tracks show a weak increase in $V-K_{\rm S}$ colour with increasing MLR and reproduce the colours reasonably well. Nevertheless, they are redder than most of the stars with Tc observations, that is, the stars are hotter and/or have more transparent dust envelopes than the models. Comparing the tracks with different ${\rm C}/{\rm O}$ ratios, we see that those with ${\rm C}/{\rm O}=0.95$ have slightly bluer $V-K_{\rm S}$ colours, indicating more transparent envelopes.

\begin{figure}
\centering
\includegraphics[width=\linewidth,bb=26 12 490 346]{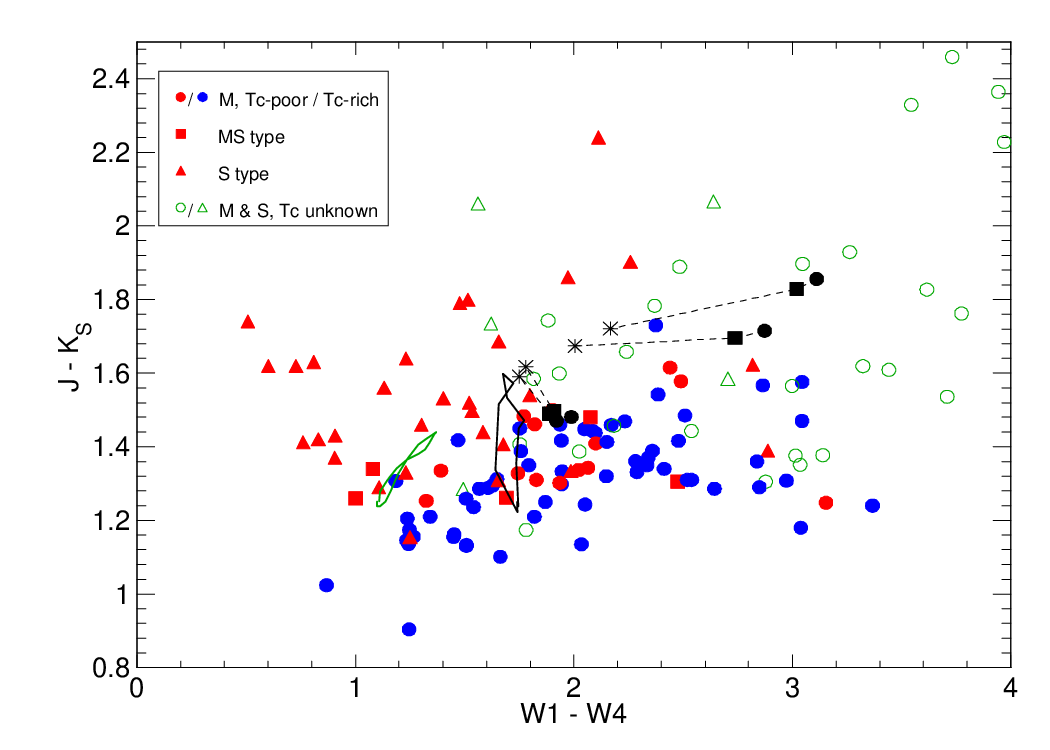}
\caption{$J-K_{\rm S}$ vs.\ $W1-W4$ colour--colour diagram with symbols as in Fig.~\ref{Fig:DUSTY1}. The figure merges the samples shown in Figs.~\ref{Fig:W1-W4_vs_P} and \ref{Fig:DUSTY1}. Dashed lines connect the \textsc{dusty}-based stationary wind models with the same MLR, increasing from left to right. The green and black lines are the DARWIN models An114u4 and M2n315u6 from Fig.~\ref{Fig:W1-W4_vs_P} (Bn114u4 almost coincides with An114u4). 
}
\label{Fig:DUSTY2}
\end{figure}

Finally, we compare the modelled and observed $J-K_{\rm S}$ vs.\ $W1-W4$ colour-colour diagram in Fig.~\ref{Fig:DUSTY2}. This figure not only includes the faint Miras shown in Fig.~\ref{Fig:W1-W4_vs_P}, but also stars with MLR determinations from Fig.~\ref{Fig:DUSTY1}. As many of these stars are brighter than the saturation limit of the WISE All-Sky Data Release in $W1$, we resorted to the unWISE catalogue \citep{2019ApJS..240...30S} to include the latter in the analysis. 
Their $W1$ photometry might not be as accurate as those of the faint Miras; nevertheless, it appears to be sufficiently accurate for our purposes and we did not find signs of saturation. A similar dichotomy as in the $W1-W4$ vs.\ $P$ diagram (Fig.~\ref{Fig:W1-W4_vs_P}) emerges as the Tc-rich stars are redder in $J-K_{\rm S}$ than the Tc-poor stars. This is most likely a result of the ${\rm C}/{\rm O}$ ratio being enhanced in the Tc-rich stars due to 3DUP episodes. \citet[][their Fig.~11]{2016MNRAS.457.3611A} demonstrated from hydrostatic models that, below about $T_{\rm eff}=3000$\,K, the $J-K_{\rm S}$ colour increases with increasing C/O. Finally, we see that the stars with unknown Tc content (but known mass-loss rate) extend the sample towards redder $J-K_{\rm S}$ and $W1-W4$ colours.

The \textsc{dusty} stationary wind models show three important effects in Fig.~\ref{Fig:DUSTY2}. First, both the $J-K_{\rm S}$ and $W1-W4$ colours increase with increasing MLR, as may be expected. The observed stars also indicate this trend. Second, at a low MLR, the $J-K_{\rm S}$ colour increases with increasing ${\rm C}/{\rm O}$, as discussed above. Third, at all mass-loss rates, the $W1-W4$ colour decreases with increasing ${\rm C}/{\rm O}$, possibly because of the amount of silicate dust produced being reduced due to the paucity of oxygen in the circumstellar envelope. This essentially confirms the results from the observational data presented in Fig.~\ref{Fig:W1-W4_vs_P}. The colour shift due to increasing ${\rm C}/{\rm O}$ is almost perpendicular to the shift in MLR, confirming the fundamental change in dust properties and/or amount with ${\rm C}/{\rm O}$ \citep[e.g.][their Fig.\ 10]{2012MNRAS.427.2647M}. In general, the stationary wind models trace the observed distribution very well, reaching the $W1-W4$ colours of the reddest Tc-poor objects, while also matching the other mass-loss parameters. 

The $W1-W4$ colours of the stationary models are all redder than those of the DARWIN models. We note, however, that the range of available DARWIN models with Fe-bearing silicate grains is very limited at present, and the properties of the stationary wind models can be steered more easily by tuning their parameters.

\subsubsection{Average spectral energy distributions of Tc-poor and Tc-rich Miras}\label{sec:Average_SED}

Compared to individual colours, a more complete picture of the emission from different types of Miras can be obtained by inspecting their SEDs. To minimise selection effects, instead of inspecting the SEDs of stars one by one, we can construct average SEDs of groups of Miras and analyse them for differences. For this aim, we collected photometry in 19 different bands that cover the range $0.3 - 90\,\mu{\rm m}$, that is in the $B$, $V$, $I$ (all from AAVSO), $J$, $H$, $K_{\rm S}$, $L$, and $M$ bands \citep{1979SAAOC...1...61C,1992A&AS...93..151F,1994MNRAS.267..711W,2000MNRAS.319..728W,2006MNRAS.369..751W,2008MNRAS.386..313W,1994A&AS..106..397K,1995A&AS..113..441K,1996A&A...311..273K,2006AJ....131.1163S,2010ApJS..190..203P}, in the $W1$ to $W4$ bands of the WISE space observatory \citep{2012wise.rept....1C,2019ApJS..240...30S}, IRAS 12, 25, and 60\,$\mu{\rm m}$ \citep{1984ApJ...278L...1N}, and Akari 9, 18, 65, and 90\,$\mu{\rm m}$ bands \citep{2010A&A...514A...1I}. To avoid signatures of stellar variability as much as possible, time-averaged photometry was adopted where available. The photometry was dereddened using the 3D map of \citet{2017AstL...43..472G}, applying the extinction law of \citet{2016ApJS..224...23X} and adopting stellar distances based on the Gaia DR3 catalogue. The extinction coefficients for some of the bands that were not determined by \citet{2016ApJS..224...23X} were interpolated from their Fig.~18, whereas no extinction correction was applied to the IRAS~60, Akari~65, and Akari~90\,$\mu{\rm m}$ bands. Since the map of \citet{2017AstL...43..472G} extends only out to 1200\,pc from the sun, extinction was set to this outer boundary for the more distant stars. 
Some of the photometric data points are missing for some of the stars due to saturation or non-detection. Each individual SED was normalised to a maximum flux of 1.0. Eventually, we calculated average fluxes for both groups of Miras in each of the photometric bands.

\begin{figure}
\centering
\includegraphics[width=\linewidth,bb=20 10 492 342]{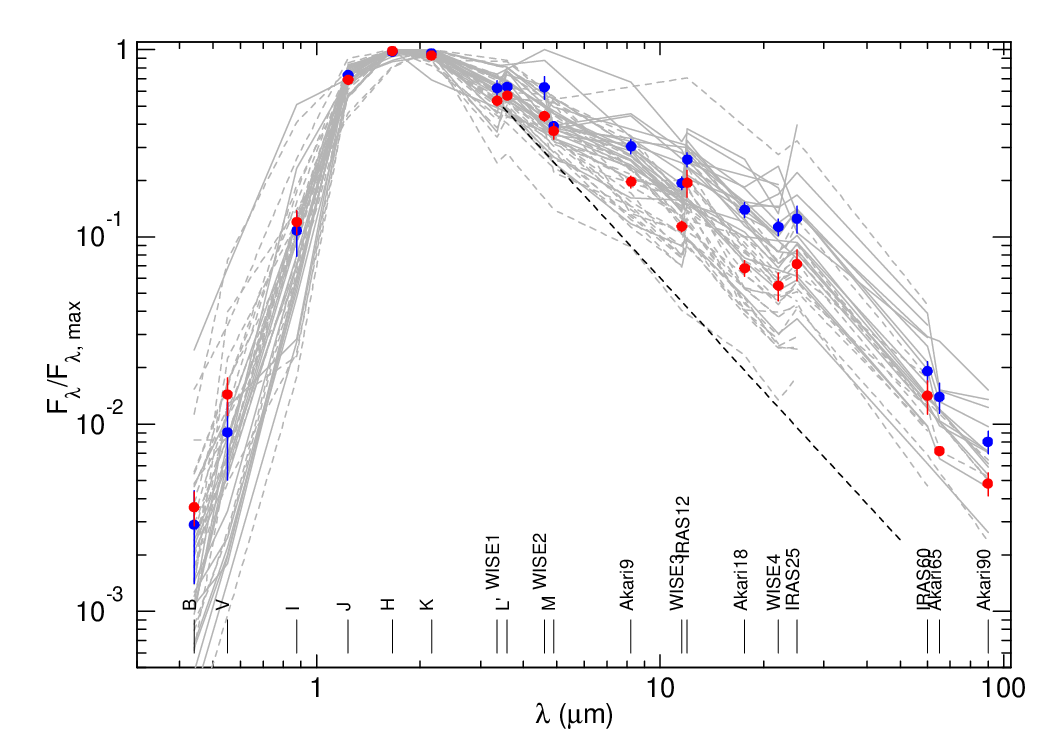}
\caption{Average SEDs of 21 Tc-poor and 26 Tc-rich Miras in the period range 280 -- 350\,d. Individual SEDs are plotted in grey in the background, where solid lines represent SEDs of Tc-poor Miras and dashed lines Tc-rich ones. Their average SEDs are represented by blue (Tc-poor) and red (Tc-rich) circles, the vertical bars indicate the standard deviation of the mean in each photometric band. The photometric bands and their central wavelengths are indicated at the bottom of the figure. The black dashed line indicates a $\lambda^{-2}$ law extending from the $W1$ band to indicate the flux of a dust-free stellar photosphere. As can be seen, the Tc-poor Miras are brighter throughout the IR range than the Tc-rich ones.}
\label{Fig:Mean_SEDs}
\end{figure}

For this comparison, we selected Miras in the period range 280 to 350\,d. The split into two sequences in the $K-W4$ colour is visible in the period range 250 to 400\,d, but because $K-W4$ increases with period in both groups, a narrower period range has to be chosen to avoid blurring of differences between the groups. With these selection criteria, we obtain 21 Tc-poor Miras and 26 Tc-rich Miras. They have mean periods of $315.4\pm16.4$ and $319.9\pm20.8$\,d ($1\sigma$ standard deviation). The result of this exercise is presented in Fig.~\ref{Fig:Mean_SEDs}. In this figure, the blue symbols represent the average SED of the Tc-poor Miras, and the red symbols those of the Tc-rich Miras. The vertical bars on the markers indicate the standard deviation of the mean in each photometric band.

On average, both groups reach their maximum flux in the $H$-band, though the $K_{\rm S}$-band flux is nearly as high. Figure~\ref{Fig:Mean_SEDs} suggests that in the optical $B$-, $V$-, and $I$-bands, the Tc-rich Miras have, on average, somewhat higher flux than the Tc-poor ones. This agrees with the result of the stationary wind models with ${\rm C}/{\rm O}=0.95$ having slightly bluer $V-K_{\rm S}$ colours than those at lower ${\rm C}/{\rm O}$ ratios, cf.\ Fig~\ref{Fig:DUSTY1}. However, the relative flux range is considerable, as can be read off from the uncertainties of the means in these bands. Looking at the long-wavelength part of the SEDs, we first notice that the IRAS bands have systematically higher fluxes than the adjacent WISE and Akari bands (note the band markers at the bottom of the figure). This could be caused by the wider IRAS filters bringing in light from shorter wavelengths \citep{2017MNRAS.471..770M} and, in areas of high background, less reliable background subtraction compared to the WISE and Akari missions due to the lower angular resolution. 
Independently of these jumps in flux, Fig.~\ref{Fig:Mean_SEDs} demonstrates that, longward from $\sim8\,\mu{\rm m}$ (Akari~9), Tc-poor Miras have significantly and consistently higher flux in the MIR than Tc-rich ones. In the $W4$ and Akari~18 bands, the average flux from the Tc-poor Miras is higher than that from Tc-rich ones by more than a factor of two. The differences in $V-K$, $K-W4$, and $W1-W4$ between the two groups of Miras (Figs.~\ref{Fig:K-W4_vs_P} and \ref{Fig:W1-W4_vs_P}) therefore reflect a change to the broader SEDs of the stars, with the Tc-rich stars having warmer (or less obscured) SEDs and less excess infrared flux above the $F_\lambda \propto \lambda^{-2}$ photosphere. 

We point out that the Tc-rich outlier at high fluxes at $\lambda\gtrsim10\,\mu{\rm m}$ (dashed line) is $o$~Ceti, which is a well-known binary system. It has already been pointed out in paper~II that Tc-rich Miras in binaries have much higher MIR fluxes than their Tc-rich siblings at similar periods without a known companion. This suggests that binaries host a large reservoir of dust in their circumstellar environment.

In summary, Fig.~\ref{Fig:Mean_SEDs} provides strong additional evidence that, at a given pulsation period, the dust emission of the CSEs of Tc-poor and Tc-rich Miras differ significantly, in the sense that Tc-poor Miras have much stronger MIR dust emission than their post-3DUP siblings. 

\section{Discussion}\label{sec:Discussion}

The data collected for this paper suggest that, at similar pulsation periods, Tc-poor Miras have i) the same gas MLRs; ii) higher terminal expansion velocity; iii) slightly redder colours in the optical spectral range; and iv) clearly higher MIR emission than Tc-rich Miras. We now discuss the evidence in light of state-of-the-art wind models.

\subsection{Fe-enrichment of dust grains}\label{sec:Fe-enrich}

The strongest observational evidence is that Tc-poor Miras have higher MIR emission than Tc-rich ones, on average by a factor of two. If the MLR is the same for both groups of Miras, then this can be interpreted by either a higher amount of dust (i.e.\ a higher dust MLR or a lower gas-to-dust ratio $\delta$) or a higher emissivity of the dust in the CSEs of Tc-poor Miras. Grains emit more MIR photons if they are hotter but, as demonstrated by \citet{2015A&A...575A.105B} and \citet{2022A&A...657A.109H}, the MIR emissivity can be greatly enhanced if the silicate grains are enriched in Fe (i.e.\ Mg$_2$SiO$_4$ $\rightarrow$ Fe$_2$SiO$_4$) or other transition metals at a level of a few per cent. The metal enrichment enhances the absorption of light, thereby increasing the temperatures of the grains, and the absorbed energy is re-emitted in the MIR. Composite grains with an Al$_2$O$_3$ core and a silicate mantle, possibly with transition-metal impurities, have been suggested by \citet{2016A&A...594A.108H} and \citet{2016A&A...585A...6G}. Such grains are still very transparent in the optical and near-IR, as required by observational evidence \citep{2016A&A...594A.108H} and also suggested by the mild difference in the optical spectral range (cf. Fig.~\ref{Fig:Mean_SEDs}). Thus, if the silicate grains in the CSEs of Tc-poor Miras have higher Fe-enrichment than those around Tc-rich Miras, they could well reproduce the observations. 

The chemical or physical mechanism at work on the microscopic level that could reduce the Fe enrichment of the grains in the CSEs of Tc-rich stars is unknown. However, some insight can come from observations of metal-poor stars. When comparing different spectral types, observations of nearby dwarf galaxies, S-type stars similarly have bluer infrared colours on average than either M-type or C-type stars \citep{2012MNRAS.427.2647M,2015ApJ...810..116B}, consistent with Fig.~\ref{Fig:DUSTY2}.

Focusing on M-type AGB stars, some stars in globular clusters, the Galactic halo and Local Group dwarf galaxies show red $K-W4$ and other infrared colours \citep{2009ApJ...705..746B,2017ApJ...851..152B,2009MNRAS.394..831M} but their spectra lack silicate features \citep{2006ApJ...653L.145L,2010ApJ...719.1274S}, presenting instead a featureless infrared excess \citep{2011ApJ...730...71M,2011MNRAS.417...20M}. This featureless excess is characteristic in shape and is best explained by metallic (predominantly iron) dust around the star \citep{2010ApJ...717L..92M}. The low mass of these stars means they have inefficient 3DUP, so should be Tc-poor. The characteristic shape of this featureless excess is very similar to the differences between Tc-rich and Tc-poor stars in Fig.~\ref{Fig:Mean_SEDs}, possibly indicating increased iron content in dust grains of Tc-poor stars, either as purely metallic grains or inclusions, or as iron doping of magnesium-rich silicate grains.

%

In the few cases where submm CO observations of metal-poor stars exist, they typically indicate a much lower outflow velocity \citep{2016A&A...596A..50G,2019MNRAS.484L..85M,2020MNRAS.491.1174M} than similar Galactic stars, though this should not be confused with a low velocity because they are Tc-poor, as submm observations of metal-poor, Tc-rich stars do not exist for comparison. Rather, the lower outflow velocity results from them being metal-poor in general, thus having a higher dust-to-gas ratio. Fe-enhancement is therefore still consistent with the faster expansion velocity of Tc-poor winds: at similar mass-loss rates, Fe-enrichment of the grains would increase the extinction cross-section, leading to more efficient momentum transfer from stellar radiation to grains, accelerating the wind.

%

\subsection{Observability of Tc lines}\label{sec:Tc-observability}

Hypothesis no.~5 presented in paper~II proposes that atomic Tc could be invisible in stars, not because it is under-abundant, but because the Tc lines are not excited or not detectable. The reasons for this could be that Tc is either ionised, incorporated into molecules or dust under certain atmospheric conditions, or Tc lines are blanketed by molecular absorption, thereby diminishing the atomic Tc absorption lines below the detection limit. Under this scenario, Tc would actually be present in more (or all) Miras, but be detectable only in those with atmospheric conditions favourable for atomic, neutral Tc to be present. This could mimic a separation of Tc-poor and Tc-rich Miras in the relevant diagrams.

We tested this hypothesis in more depth using the COMARCS model atmospheres of \citet{2016MNRAS.457.3611A}. We determined the temperature of the layers in which the Tc lines are mainly formed with the cut-off method, in other words, cutting off atmospheric layers with temperatures lower than a certain threshold. Depending on the effective temperature of the model and the adopted ${\rm C}/{\rm O}$ and $[{\rm Zr}/{\rm H}]$ enhancements, the lines are formed in the temperature range $1500-2500$\,K. \ion{Tc}{I} and \ion{Tc}{II} exist in equal amounts at temperatures of 3500\,K or higher, slightly depending on the density in the atmosphere. Thus, Tc gets significantly ionised only in layers that are much hotter than the Tc line-formation zone. We can also exclude shocks to ionise Tc because they would also destroy ZrO and other molecules, which are clearly observed in the spectra. Our tests with synthetic spectra also show that veiling by other line opacities, in particular molecular opacity, may slightly reduce the Tc line strength, but is unable to hide them completely. We note that stars with prominent ZrO bands all have observable Tc, speaking against veiling having a significant effect. It is also not probable that emission in the extended dynamic atmosphere fills in the absorption lines because different Tc lines show the same trend. Only Mg, Mn, and Fe lines are known to appear in emission sometimes \citep{1962ApJ...136...21M}. Many stars have been searched for Tc lines more than once, and the results are generally consistent for observations collected at different pulsation phases; see, for example, the observations collected in paper~I and II. One exception could be R~Hya, for which conflicting results have been reported: It was classified as Tc-poor based on some spectra, but other observations demonstrated that it does have Tc lines, albeit relatively weak; see the discussion in \citet{2011A&A...531A..88U}. Thus, a few stars that actually underwent 3DUP and have Tc in their atmospheres could erroneously be classified as Tc-poor.
%

There are additional arguments as to why this hypothesis could not be applicable. One is that a high degree of fine-tuning to obtain a separation between Tc-poor and Tc-rich Miras as clean as in Fig.~\ref{Fig:W1-W4_vs_P} would be required. As demonstrated in Sect.~\ref{sec:ZrO}, atomic Tc lines appear in the stellar spectra in lockstep with molecular ZrO lines. Even though the ZrO bands increase in strength, the atomic Zr lines (and those of other s-elements) do not become weaker, quite the opposite. Thus, Tc would have to be ionised or incorporated in the molecules very selectively, without touching the other $s$-elements.

Even though this hypothesis is not likely to explain the observations, we qualitatively explore in the following the possibility that Tc could be selectively removed from the gas phase in O- and C-rich AGB atmospheres into nanometer-sized clusters and dust grains. In the case of O-rich environments, Tc could be present as solid technetium (IV) oxide, TcO$_2$, or technetium (VII) oxide, Tc$_2$O$_7$. Whereas TcO$_2$ is refractory with a melting point of 1370\,K, Tc$_2$O$_7$ condenses at comparatively low temperatures of 393\,K \citep{Nelson1954}. Therefore, solid Tc$_2$O$_7$ is not expected to be present in the inner atmospheric layers of the hot circumstellar envelopes, but the existence of TcO$_2$ condensates seems to be more promising. Tc-bearing gas phase species, including its related clusters that represent potential molecular building blocks of solid technetium oxides, could exist in stellar atmospheres and are a potential target for future observational studies. The thermal stability of these clusters depends on many factors including stoichiometry, size, geometry, and spectroscopic constants, which can differ substantially from the corresponding technetium oxide condensates. Moreover, the reaction kinetics of the formation, oxidation, and growth of Tc-bearing clusters are likely to involve barriers and could even prevent the formation of a particularly stable molecule or cluster. However, due to the lack of laboratory and theoretical data, the properties of nanoscale Tc oxides and their related formation and growth rates are not known. Therefore, the presence and abundance of Tc-containing clusters are challenging to predict, and its derivation based on the thermal stability of the bulk materials would only fall too short. In carbon-dominated chemistry, carbonaceous Tc compounds could exist. The existence of crystalline technetium monocarbide (TcC) has been predicted \citep{GIORGI1966455}. However, a recent theoretical study has shown that solid TcC does not exist and that the previously reported TcC crystal structure corresponds to a high-temperature cubic phase of elemental technetium \citep{C5RA24656C}. Nevertheless, this finding does not rule out the possible existence of carbonaceous Tc-bearing molecules and clusters. We aim to address the possibility of Tc-containing molecules, clusters, and condensates in more detail in a forthcoming study (Gobrecht et al., in preparation).

\subsection{Reduction of free oxygen by dredge-up of $^{12}$C}\label{sec:O-lack}

Repeated dredge-up of $^{12}$C will increase the atmospheric C/O ratio and bind more oxygen in CO, thus reducing free O available to form molecules and dust. Silicates (e.g.\ Mg$_{2x}$Fe$_{2(1-x)}$SiO$_4$) and alumina (Al$_2$O$_3$) require large amounts of O to form, and thus their formation could be hampered at an increased C/O. Oxygen-free dust grains (e.g.\ metallic iron or MgS) are discussed as alternative wind drivers at ${\rm C}/{\rm O}\approx1$ \citep[e.g.][]{2012A&A...540A..72S}.

We can make a simple estimate at which C/O ratio free oxygen will become the bottleneck for the formation of silicate dust species such as Mg$_2$SiO$_4$ \citep[cf., discussion in][]{2011ApJ...730...71M}. In the solar composition \citep[e.g.][]{2003ApJ...591.1220L}, Mg and Si have essentially the same abundance, namely $N({\rm Mg})=1.020\times10^6$ on a scale where $N({\rm Si})\equiv10^6$. Carbon ($N({\rm C})=7.079\times10^6$) and oxygen ($N({\rm O})=1.413\times10^7$) are much more abundant, but almost all C and the same amount of O are locked up in CO. Hence, only $O-C=7.051\times10^6$ of free oxygen is available to form dust. Some O will actually form MgO and SiO molecules. However, four times more oxygen than Si atoms are required to form Mg$_2$SiO$_4$, so O must be freely available in four times the amount as Mg and Si to not hinder the formation of the cluster, i.e.\ ${\rm O}-{\rm C}\geq4\times10^6$. Assuming the O abundance to be constant at $N({\rm O})=1.413\times10^7$ and the C abundance to be increased by 3DUP, this translates into a limit of ${\rm C}/{\rm O}\gtrsim0.72$ above which free oxygen will be the bottleneck for the formation of Mg$_2$SiO$_4$ dust grains. AGB evolutionary models predict this C/O ratio will be reached after only a few 3DUP events.

The O-bearing molecules H$_2$O and, in particular, OH are essential to the dust-nucleation processes, leading to the formation of circumstellar dust grains \citep[e.g.][]{2022A&A...658A.167G}. Submillimetre observations of O-rich AGB stars reveal that the H$_2$O abundance is reduced in S-stars, which have an enhanced C/O ratio compared to M-stars \citep[e.g.][]{2014A&A...569A..76D}. In the sample of \citet{2023A&A...674A.125B}, the only two targets without any H$_2$O transition detection are the two S-type stars. Chemical models such as those of \citet{2020A&A...637A..59A} corroborate these observational results.

A decrease in the dust-to-gas ratio as stars undergo 3DUP and free oxygen is reduced may well explain all our observations. \citet{2016ApJ...823L..38M} and \citet{2019MNRAS.484.4678M} showed that the pulsation properties of stars that set their MLR, with pulsation amplitude being the most important. Our Fig.~\ref{Fig:Mdotgas_vs_P} emphasises this: there is a strong link between gas MLR and pulsation period (and amplitude) that does not differ between the two sets of stars. The MLR does not change with 3DUP, but the wind is still dust-driven, and the dust-to-gas ratio falls. This reduces the optical depth of dust in the SED and the reprocessing of light, decreasing the amount of infrared emission while recovering some of the optical flux (Fig.~\ref{Fig:Mean_SEDs}). The wind cannot be ejected as efficiently by dust driving, so the expansion velocity falls (Fig.~\ref{Fig:vexp_vs_P}). Our stationary wind models (Sect.~\ref{sec:DUSTY}) confirm this reasoning and reproduce the observations: The MIR emission drops with increasing C/O, reflected by a decrease in $W1-W4$ colour (Fig.~\ref{Fig:DUSTY2}, while the $V-K_{\rm S}$ colour becomes only marginally bluer (Fig.~\ref{Fig:DUSTY1}), lower panel). At the same time, the expansion velocity drops (Fig.~\ref{Fig:DUSTY1}, upper panel).

\begin{figure}
\centering
\includegraphics[width=\linewidth,bb=26 10 488 344]{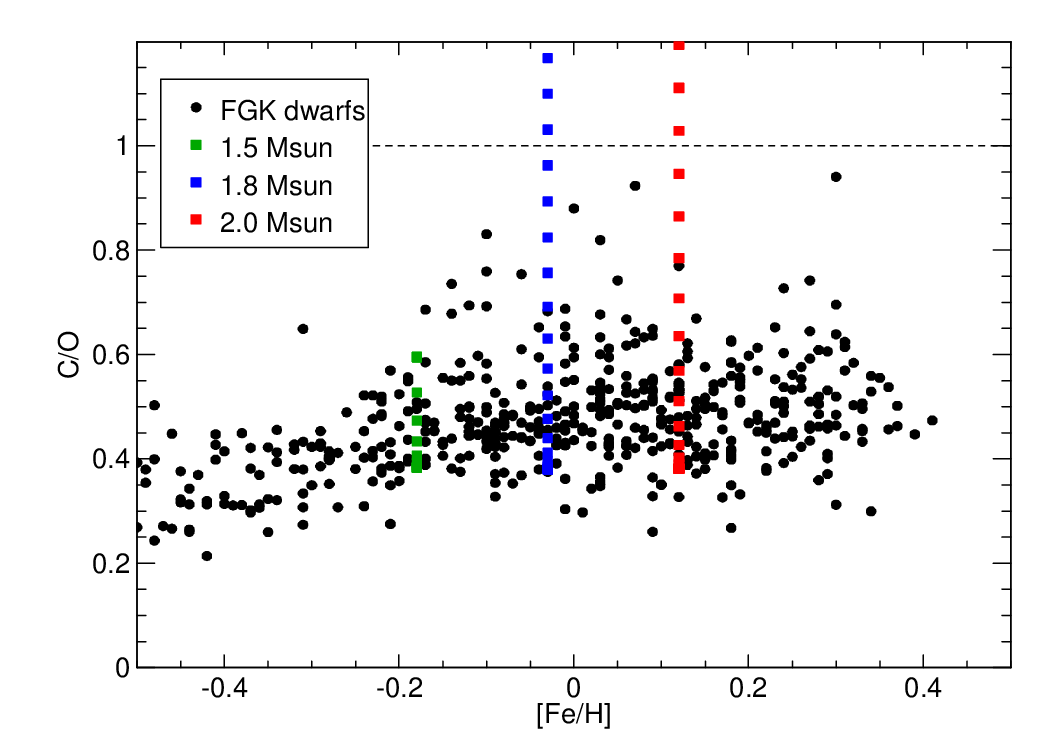}
\caption{C/O vs.\ [Fe/H] of FGK dwarf stars from \citet{2021A&A...655A..99D}. Also plotted is the evolution of C/O of three \texttt{COLIBRI} TP-AGB models from \cite{Marigo_etal13} with masses of 1.5, 1.8, and 2.0\,$M_{\sun}$. All models are calculated at $[{\rm Fe}/{\rm H}]=-0.03$, but the models with 1.5 and 2.0\,$M_{\sun}$ are shifted in [Fe/H] for clarity. Each symbol on the TP-AGB tracks marks a 3DUP episode. The horizontal dashed line marks ${\rm C}/{\rm O}=1$, above which the 1.8 and 2.0\,$M_{\sun}$ models become carbon-rich.}
\label{Fig:COdwarfs}
\end{figure}

An interesting test of this 'lack-of-oxygen hypothesis' is to inspect how large the C/O increase by one or few 3DUP episodes is compared to the intrinsic spread in C/O ratios of stars. If the stars' intrinsic C/O spread when they enter the AGB phase was large compared to the C/O increase by 3DUP events, the Tc-poor stars alone would exhibit a significant intrinsic variation of the IR dust emission. Precise determinations of C/O in pre-3DUP AGB stars are difficult and rare, but observations of main-sequence stars are readily available. For example, \citet{2021A&A...655A..99D} determined [C/O] for a sizeable sample of FGK stars, which we transform to C/O ratios using their adopted solar value ${\rm C}/{\rm O}_{\sun}=0.55$ (see their Sect.~6.1). These C/O abundances will be changed (reduced) by the first dredge-up once main-sequence stars evolve to the RGB, but we may assume that this event will not wipe out the intrinsic, primordial C/O spread with which they will enter the AGB. A zoom-in on metallicities close to solar in the C/O vs.\ [Fe/H] plane of the \citet{2021A&A...655A..99D} sample is shown in Fig.~\ref{Fig:COdwarfs}. Most objects in this metallicity range are thin-disc members. Here, we adopted the more precise measurements based on the \ion{O}{I} 6158\,\AA\ line. Over-plotted are three TP-AGB model tracks computed with the \texttt{COLIBRI} code of \cite{Marigo_etal13} with masses 1.5, 1.8, and 2.0\,$M_{\sun}$. Each symbol represents the C/O value after a 3DUP episode.

%

Dwarf stars in the sample of \citet{2021A&A...655A..99D} with solar-like metallicity ($[{\rm Fe}/{\rm H}]=\pm0.20$, 310 stars) have a C/O standard deviation of $\sim0.10$. This is actually less than the average formal uncertainty quoted by \citet{2021A&A...655A..99D} of 0.11. Thus, the formal uncertainty slightly over-predicts the width of the C/O distribution of solar-like stars. We must therefore assume that the intrinsic spread in C/O could be much smaller than 0.1. Stars entering the AGB phase should thus have a relatively narrow range in C/O before they experience 3DUP. \citet[][their Fig.~15]{2000A&A...362..599G} estimated how many TPs with subsequent 3DUP events are required for Tc to become detectable on the stellar surface. With their 2.5\,$M_{\sun}$, $Z=0.008$ model they estimate that Tc is detectable on the surface after five TPs. The \texttt{COLIBRI} evolutionary models with 1.8 and 2.0\,$M_{\sun}$ predict that seven and six 3DUP events, respectively, are required to enhance C/O by more than $1\sigma$ of the observed C/O distribution (note that the TP-AGB models start out with the C/O value left after 1DUP, which is slightly lower than on the main sequence due to partial CNO cycling). The value of ${\rm C}/{\rm O}\gtrsim0.72$, above estimated to be the threshold when free oxygen becomes the limiting factor in the formation of silicate dust grains, is reached after twelve and nine 3DUP events, respectively, in these models. Once Tc is detectable, the surface C/O should thus have reached a level where the available amount of oxygen becomes a limiting factor for dust formation.

State-of-the-art DARWIN and \textsc{dusty} stationary wind models do not include a full modelling of dust formation including primary nucleation but assume the seed particle abundance as a free parameter. Therefore, the effect of C/O enhancement on the formation of the first condensation nuclei cannot yet be tested with the wind models. This point should be addressed in future work.

\subsection{Further considerations}\label{sec:Further-cons}

We also need to consider that the current data quality (and the number of observed stars) may still be too limited to rule out a difference in gas MLR between Tc-poor and Tc-rich Miras. The currently ongoing large programmes ATOMIUM \citep{2022A&A...660A..94G}, DEATHSTAR \citep{2020A&A...640A.133R}, and NESS \citep{2022MNRAS.512.1091S} aim at improving the MLR measurements by CO observations with modern observatories and radiative transfer models. Determining $\dot{M}_{\rm gas}$ from the mm-wave CO line observations depends on a number of assumptions \citep[CO/H$_2$ ratio, CO envelope size, sphericity, constant mass-loss rate, etc.; e.g.\ ][]{2021A&A...653A..53A} and further uncertain parameters of the stars such as their distance \citep[thus luminosity; e.g.\ ][]{2022A&A...667A..74A}, the interstellar radiation field, etc. Precise distances for many AGB stars have now become available with the Gaia catalogues. We can expect that the data quality may improve significantly in the near future, but both distance and detection of Tc remain unmeasured for optically enshrouded stars. Thus, once gas MLRs become available from these programmes, the points addressed in this paper should be revisited.

On the other hand, the expansion velocity $v_{\infty}$ of the CSE can be measured much more directly with fewer assumptions from the observations than the gas MLR. This might reduce scatter in $v_{\infty}$ and could explain why we see a difference between Tc-poor and Tc-rich Miras in $v_{\infty}$, but not in $\dot{M}_{\rm gas}$. Models and observations show a well-defined relation between $\dot{M}_{\rm gas}$ and $v_{\infty}$ \citep[e.g.][]{2009A&A...499..515R}. One would thus not necessarily expect the two groups of Miras to distribute differently in the $\dot{M}_{\rm gas}$ versus\ $P$ plane compared to $v_{\infty}$ versus\ $P$.

Another factor that might compromise detecting a difference between the two groups of Miras in terms of gas MLR is the fact that colours such as $K-W4$ probe the ML on a different timescale than the CO mm-wave lines. While $K-W4$ probes the warm dust, that is, ML from the last few decades, the CO lines probe the ML history back to $\sim15\,000$\,yrs \citep{2019IAUS..343..150R}. The MLR estimates based on rotational CO line observations are representative of the average MLR over the time of the creation of the emitting region and are not necessarily a good measure of the current MLR from a star. However, the pulsation period, which is the stellar property that we compare against in the diagrams, probes the current state of the stellar atmosphere. It may change considerably even over a timescale of decades, for example, by a TP going on in the deep interior \citep{2011A&A...531A..88U} or due to chaotic changes and feedback \citep{1996ApJ...456..350Y,2004MNRAS.352..325Z}. Thus, the period might not be so directly related to the MLR measured from CO lines as to the $K-W4$ colour. On the grand scheme, the gas MLR measured from CO and the present-day pulsation periods show a clear correlation (e.g. Fig.~19 of \citet{2018A&ARv..26....1H} and Fig.~8 of paper~II), but any finer detail of the correlation might be more difficult to reveal due to scatter of the data.


Finally, we also mention again the evidence for the impact of binarity on the ML process. Compared to other Tc-rich Miras with similar pulsation periods, known and suspected binaries have an enhanced $K-W4$ colour (Fig.~\ref{Fig:K-W4_vs_P}) and $v_{\infty}$ (Fig.~\ref{Fig:vexp_vs_P}). Both these pieces of evidence support recent theoretical results such as those of \citet{2021A&A...653A..25M}. There is no known binary among the Tc-poor Miras presented in this paper, so we cannot make any statements about the impact of binarity in this group. However, the short-period Mira R~Cet is very red in $K-W4$ and has a comparably high terminal expansion velocity. Thus, we propose it as a binary candidate that should be considered a target for future mm-wave CO observations.

\section{Conclusions}\label{sec:conclusio}

Our increased sample does not show a clear distinction between Tc-poor and Tc-rich Miras in the gas mass-loss rate as a function of the pulsation period. On the other hand, the expansion velocity appears to be higher in Tc-poor Miras compared to Tc-rich ones at similar pulsation periods. We provide further evidence that sequences of Miras emerge in a NIR-to-MIR-colour versus\ pulsation period diagram if a distinction is made for the presence of Tc in the stars. In particular, the two groups of Miras are clearly separated in the $W1-W4$ versus\ period diagram (Fig.~\ref{Fig:W1-W4_vs_P}). Furthermore, average SEDs confirm earlier results that Tc-poor and Tc-rich Miras differ in their amount of MIR emission: Tc-poor Miras emit more MIR light than Tc-rich ones by up to a factor of two relative to the SED maximum (Fig.~\ref{Fig:Mean_SEDs}).

We suggest that the observations can be explained by repeated 3DUP events that will reduce the availability of free oxygen for the formation of silicate dust. Tc lines are an indicator of 3DUP going on in a star before other spectral features are noticeably affected by this deep mixing event. Our observations suggest that the dust formation and mass-loss process are also already affected by a few 3DUP events. Another possibility is that the dust grains formed around Tc-poor, pre-3DUP stars are more enriched in iron and other transition metals and thus reprocess more stellar light to the MIR than Tc-rich stars. This corresponds to hypothesis 4 introduced in paper~II, which we discuss in more depth here (Sect~\ref{sec:Fe-enrich}). Following from the discussion in Sect.~\ref{sec:Tc-observability}, on the other hand, we drop the hypothesis of selective Tc line observability to explain the observations (hypothesis 5 in paper~II).

Dedicated chemical and dust-formation models in AGB envelopes are required to better understand the impact of 3DUP on the mass loss from AGB stars. On the observational side, with results being made available from the ongoing large programmes ATOMIUM, DEATHSTAR, and NESS, the aim of which is to reduce the uncertainties in the mass-loss rate determination from AGB stars, it will be useful to again revisit the question of the impact of 3DUP on the mass-loss rate of Miras.


\begin{acknowledgements}
Based on data obtained with the TIGRE telescope, located at La Luz observatory, Mexico. TIGRE is a collaboration of the Hamburger Sternwarte, the Universities of Hamburg, Guanajuato and Li\`{e}ge. Based on observations made with the Mercator Telescope, operated on the island of La Palma by the Flemish Community, at the Spanish Observatorio del Roque de los Muchachos of the Instituto de Astrof\'{i}sica de Canarias. We acknowledge with thanks the variable star observations from the AAVSO International Database contributed by observers worldwide and used in this research. This publication makes use of data products from the Two Micron All Sky Survey, which is a joint project of the University of Massachusetts and the Infrared Processing and Analysis Center/California Institute of Technology, funded by the National Aeronautics and Space Administration and the National Science Foundation. This publication makes use of data products from the Wide-field Infrared Survey Explorer, which is a joint project of the University of California, Los Angeles, and the Jet Propulsion Laboratory/California Institute of Technology, funded by the National Aeronautics and Space Administration. This work has made use of data from the European Space Agency (ESA) mission Gaia (\url{https://www.cosmos.esa.int/gaia}), processed by the Gaia Data Processing and Analysis Consortium (DPAC, \url{https://www.cosmos.esa.int/web/gaia/dpac/consortium}). Funding for the DPAC has been provided by national institutions, in particular the institutions participating in the Gaia Multilateral Agreement. SS would like to acknowledge support from the Research Foundation-Flanders (grant number: 1239522N). AN acknowledges support from the Narodowe Centrum Nauki (NCN), Poland, through the SONATA BIS grant UMO-2020/38/E/ST9/00077. DG acknowledges funding from the Knut and Alice Wallenberg Foundation (research grant KAW 2020.0081). SH acknowledges funding from the European Research Council (ERC) under the European Union’s Horizon 2020 research and innovation programme (Grant agreement No. 883867, project EXWINGS) and the Swedish Research Council (Vetenskapsradet, grant number 2019-04059). The DARWIN computations were enabled by resources in project NAISS 2023/5-138] provided by the National Academic Infrastructure for Supercomputing in Sweden (NAISS) at UPPMAX, funded by the Swedish Research Council through grant agreement no. 2022-06725. SC acknowledges funding by the European Union -- NextGenerationEU RFF M4C2 1.1 PRIN 2022 project '2022RJLWHN URKA' and by INAF Theory Grant 'Understanding R-process \& Kilonovae Aspects (URKA)'.
\end{acknowledgements}

\bibliographystyle{aa}
\bibliography{ReferencesTIGRE}

\end{document}